\documentclass[final,3p,times,review]{elsarticle} 

\usepackage{amssymb,amsmath,array} 
\usepackage[colorlinks=true,citecolor=blue,linkcolor=blue,
            urlcolor=blue,pdfauthor=author]{hyperref}
\usepackage{float} 
\usepackage[flushleft]{threeparttable} 
\usepackage{multirow}
\usepackage{multicol}
\usepackage{booktabs}
\usepackage{lscape}
\usepackage{pdflscape}
\usepackage{amsmath}
\usepackage{makecell}
\usepackage{subcaption}
\usepackage{graphicx}

\usepackage{nomencl}
\makenomenclature
\usepackage{etoolbox}
\renewcommand\nomgroup[1]{%
	\item[\bfseries
	\ifstrequal{#1}{A}{Symbols}{%
		\ifstrequal{#1}{B}{Subscripts}{%
			\ifstrequal{#1}{C}{Abbreviations}{}}}%
	]}

\usepackage{soul} 

\usepackage{color} 

\usepackage{xcolor} 

\definecolor{mc1}{rgb}{0,0,0} 

\journal{Applied Energy} 

\usepackage{lineno} 

\AtBeginDocument{%
  \def\thefnote{\myfnsymbol{fnote}}}
\makeatletter
\def\myfnsymbol#1{\expandafter\@myfnsymbol\csname c@#1\endcsname}
\def\@myfnsymbol#1{\ensuremath{\ifcase#1\or \dagger\or \ddagger\or
   \mathsection\or \mathparagraph\or \|\or **\or \dagger\dagger
   \or \ddagger\ddagger \else\@ctrerr\fi}}
\def\fntext[#1]#2{\g@addto@macro\@fnotes{%
   \refstepcounter{fnote}\elsLabel{#1}%
   \def\thefootnote{\thefnote}
   \global\setcounter{footnote}{\c@fnote}%
   \footnotetext{#2}}}
\makeatother

\begin{document}

\begin{frontmatter}

\title{Efficient Estimation of Convective Cooling of Photovoltaic Arrays with Various Geometric Configurations: a Physics-Informed Machine Learning Approach}

\author[add1]{Dapeng Wang} 
\author[add1]{Zhaojian Liang} 
\author[add1]{Ziqi Zhang}
\author[add1]{Mengying Li \corref{cor1}}
\ead{mengying.li@polyu.edu.hk} 
 
\cortext[cor1]{Corresponding author} 

\address[add1]{Department of Mechanical Engineering \& Research Institute for Smart Energy, The Hong Kong Polytechnic University, Hong Kong SAR}

\begin{abstract} 

Convective heat transfer is crucial for photovoltaic (PV) systems, as the power generation of PV is sensitive to temperature. The configuration of PV arrays has a significant impact on convective heat transfer by influencing turbulent characteristics. Conventional methods of quantifying the configuration effects are either through Computational Fluid Dynamics (CFD) simulations or empirical methods, which face the challenge of either high computational demand or low accuracy, especially when complex array configurations are considered. This work introduces a novel methodology to quantify the impact of geometric configurations of PV arrays on their convective heat transfer rate in the wind field. The methodology combines Physics Informed Machine Learning (PIML) and Deep Convolution Neural Network (DCNN) to construct a robust PIML-DCNN model to predict convective heat transfer rates.  In addition, an innovative loss function, termed Pocket Loss, is proposed to enhance the interpretability of the PIML-DCNN model. The model exhibits promising performance, with a relative error on test dataset of 1.3\% and overall $R^2$ of 0.99 in estimating the coefficient of convective heat transfer, when compared with full CFD simulations. Therefore, the proposed model has the potential to efficiently guide the configuration design of photovoltaic arrays for the enhancement of power generation in real-world operations. 

\end{abstract}

\begin{highlights} 
\item A PIML-DCNN model is developed to estimate convective heat dissipation in photovoltaic (PV) arrays.
\item Data from 160 CFD simulations is utilized to train and validate  the PIML-DCNN model.
\item A Pocket Loss function is developed to improve the interpretability of the PIML-DCNN model.
\item A relative error of 1.9\% and overall $R^2$ of 0.99 are achieved when comparing the PIML-DCNN model against CFD results.
\item The PIML-DCNN model outperforms the empirical approach in accuracy and is computationally more efficient than direct CFD simulations.

\end{highlights}


\end{frontmatter} 
\mbox{}
\nomenclature[A]{$r_i$}{Length of gliding box, [m]}
\nomenclature[A]{$s(r)$}{Ratio of volume occupied by gliding box}
\nomenclature[A]{$\Lambda$}{Lacunarity}
\nomenclature[A]{$L_c$}{Characteristic length used to calculate Re, [m]}
\nomenclature[A]{$L,W,H$}{Length, width and height of  the computation domain, [m]}
\nomenclature[A]{$T_\infty$}{Ambient air temperature, [K]}
\nomenclature[A]{$T$}{Temperature, [K]}
\nomenclature[A]{$U_\infty$}{Wind velocity, [m/s]}
\nomenclature[A]{$T_0$}{Initial temperature, [K]}
\nomenclature[A]{Nu}{Nusselt number}
\nomenclature[A]{$k$}{Thermal conductivity of air, [W/m K]}
\nomenclature[A]{$h$}{Coefficient of convective heat transfer, [W/m$^2$.K]}
\nomenclature[A]{$D$}{Characteristic height of the array used to calculate Nu, [m]}
\nomenclature[A]{Re}{Reynolds number}
\nomenclature[A]{$\nu$}{Kinematic viscosity of air, [N s/m$^2$]}
\nomenclature[A]{Pr}{Prandlt number}
\nomenclature[A]{$S$}{Spacing between two rows of panels, [m]}
\nomenclature[A]{$\Gamma$}{Distinguish label in training, [m$^2$/s]}
\nomenclature[A]{$\Phi$}{Convolution Neural Network}
\nomenclature[A]{$\Psi$}{Geometric features}
\nomenclature[A]{$\xi$}{Pocket Loss}
\nomenclature[A]{$C_{\text{up}}$}{Upper boundary in Pocket Loss}
\nomenclature[A]{$C_{\text{low}}$}{Lower boundary in Pocket Loss}
\nomenclature[A]{$w$}{Weights of neural network}
\nomenclature[A]{$\chi$}{Parameters need physical interpretation}
\nomenclature[A]{$\epsilon$}{Relative error}
\nomenclature[A]{$e$}{Maximum relative error}
\nomenclature[A]{$R^2$}{Coefficient of determination}
\nomenclature[A]{$\beta_p$}{Temperature coefficient of photovoltaic panels, [\%/K]}
\nomenclature[A]{$P$}{Generated power from PV, [W/m$^2$]}
\nomenclature[A]{$p_0$}{Ambient air pressure,  [atm]}
\nomenclature[A]{$\alpha$}{Coefficient of Pocket Loss}
\nomenclature[A]{$d$}{Panel length, [m]}
\nomenclature[A]{$t$}{Panel thickness, [m]}
\nomenclature[A]{$\theta$}{Tilt angle of the panels, [rad]}
\nomenclature[A]{$q$}{Heat flux, [W/m$^2$]}
\nomenclature[A]{$\theta$}{Feed forward layers}
\nomenclature[A]{$r_{\text{PO}}$}{Power output ratio }

\nomenclature[B]{$i$}{$i^{th}$ element}
\nomenclature[B]{surf}{surface}
\nomenclature[B]{STC}{Standard parameters}
\nomenclature[B]{mod}{PV panel module}
\nomenclature[B]{2D}{Geometry in 2D scale}
\nomenclature[B]{3D}{Geometry in  3D scale}

\nomenclature[C]{CFD}{Computation Fluid Dynamics}
\nomenclature[C]{LES}{Large Eddy Simulation}
\nomenclature[C]{GBM}{Gliding Box Method}
\nomenclature[C]{DCNN}{Deep Convolution Neural Network}
\nomenclature[C]{PIML}{Physics-Informed Machine Learning}
\nomenclature[C]{PL}{Pocket Loss}
\nomenclature[C]{PV}{Photovoltaic}
\nomenclature[C]{RMSE}{Root Mean Square Error}
\nomenclature[C]{MAE}{Mean Absolute Error}
\nomenclature[C]{DL}{Deep Learning}

\printnomenclature

\section{Introduction} 
As one of the most popular energy conversion devices for generating power from solar energy, the photovoltaic (PV) panel is playing a crucial role in the transition toward sustainable energy sources. To ensure the efficiency and durability of PV panels, effective cooling strategies are essential to prevent excessive temperatures that affect PV performance~\cite{SIECKER2017192, DASILVA20101985, Hamid}. Recent advancements have seen a range of cooling strategies, with active and passive methods being the main categories.

\begin{figure}[htbp]
  \centering
  \includegraphics[width=0.8\textwidth]{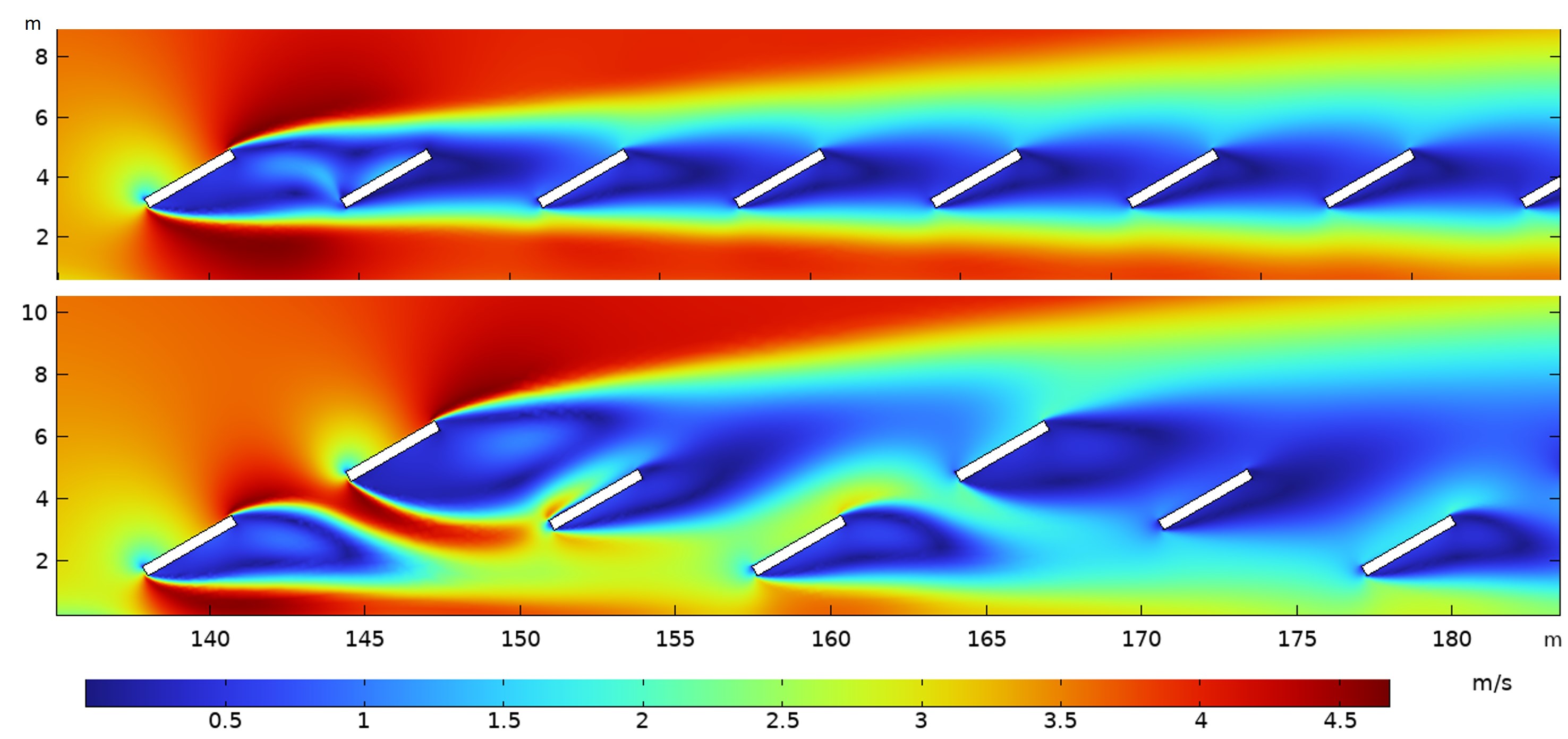}
  \caption{Air velocity over the panels of  same-height configuration (upper) and stagger-height configuration (lower). The results are generated using COMSOL. }
  \label{fig:CFDs}
\end{figure}

Active cooling, such as jet impingement, forced airflow, and liquid spraying, can effectively mitigate the efficiency losses in PV caused by high temperatures~\cite{ROYNE20071014,KAISER201488, ELNOZAHY2015100}. Some liquid-cooled systems have reported improvements in PV efficiency of up to 64\%~\cite{Rajasekar}. However, active cooling necessitates additional power input, which may offset a portion of the gains achieved through efficiency improvements~\cite{DWIVEDI2020100674}. Furthermore, in arid regions where solar resources are abundant, the scarcity of water could limit the feasibility of water-based active cooling methods. 
Passive cooling strategies, which take advantage of natural radiative and convective heat transfer, offer a more sustainable alternative by reducing the energy demands of cooling systems~\cite{DWIVEDI2020100674}. Minimizing radiative heat gain is an effective method of cooling, achievable by spectrally selective filtering of photons. This can be done through the modification of the internal structures of the cover glass or by applying specialized coatings~\cite{ZHAO2020114548,Zhu:14,LEE2019222,JEONG2020110296}. 
However, the high cost of specialized radiative coatings currently hinders their application in large-scale photovoltaic systems~\cite{AHMED2021100776}.
Another method is to enhance convective heat dissipation by arranging panels with staggered height~\cite{SMITH2022119819}, which can enhance air flow over subsequent panels, as shown in Fig.\,\ref{fig:CFDs}. 

Array configurations significantly influence convective heat transfer within PV fields. Accurate quantification of the impacts of array configurations under various environmental conditions requires extensive computational fluid dynamics (CFD) simulations and/or field experiments. 
However, these methods are resource-intensive; for example, Large Eddy Simulations (LES) can cost 43,000 core hours in a high performance computing cluster to compute 6 cases of LES 3D transient study, utilizing the Uintah MPMICE software~\cite{Stanislawski}. For fast estimation, empirical correlations~\cite{thermo3010010,BILAWANE20217926} have been developed based on CFD and/or field test data. These empirical correlations have achieved some success in estimating heat transfer for single photovoltaic panels with different orientations and heights. However, most of them assume that natural convection is the principal mechanism of heat transfer. Such assumption may not hold for predicting the heat transfer in PV arrays, where the flow fields are more complicated ~\cite{9300740}. 

To develop empirical correlations for PV arrays, some studies ~\cite{PhysRevE.53.5461,FRAZER200565,Scott_2022,smithview} introduce the concept of lacunarity, which successfully characterizes the characteristics of PV panel configurations within an array. However, the challenges in determining characteristic length scales and the limitation of the small sample size impair the accuracy and generality of the empirical correlations. 
In sum, current methodologies for quantifying convective heat dissipation in PV arrays sit at two extremes: the empirical approach, which offers computational efficiency at the cost of accuracy, and the CFD approach, which ensures accuracy but with high computational demand. 
Therefore, this paper aims to address the critical research gap by exploring an innovative approach that achieves high accuracy and computational efficiency through the application of machine learning (ML) algorithms. 

 In the field of physical problem-solving, such as modeling hydrological systems and turbulent heat transfer~\cite{BHASME2022128618,Kim_Kim_Lee_2023}, ML algorithms have recently demonstrated considerable promise for the rapid estimation of solutions. Particularly, the incorporation of physical laws into machine learning -- physics-informed machine learning (PIML) algorithms -- has significantly enhanced the interpretability of ML models when applied to complicated physical systems~\cite{Karniadakis,BHASME2022128618}. The work by \citet{ZOBEIRY2021104232}, which applies PIML to solve heat transfer equations, provide a basis for our investigation into the application of PIML for heat transfer within PV arrays. 

In this work, a new model is developed, physics-informed deep convolutional neural network (PIML--DCNN), which accurately estimates the convective heat transfer coefficient in photovoltaic arrays while keeping computational demands low. Our model uses a DCNN to process physical information from PV arrays, making it robust enough to handle different geometrical configurations and environmental conditions. In addition, we improved the model's interpretability by proposing a new loss function called `pocket loss' for PIML, which provides physical insights into the model's estimation. In the following, Section~\ref{Sec:Method} presents the methodology of developing benchmark models and the proposed PIML-DCNN model. Section~\ref{Sec:ResultsandDiscussion} presents comparative results of the considered models. Concluding remarks are summarized in Section\,\ref{sec:conclusion}.

\section{Methodology}\label{Sec:Method}
\begin{figure}[htbp]
  \centering
  \includegraphics[width = 1.0\linewidth]{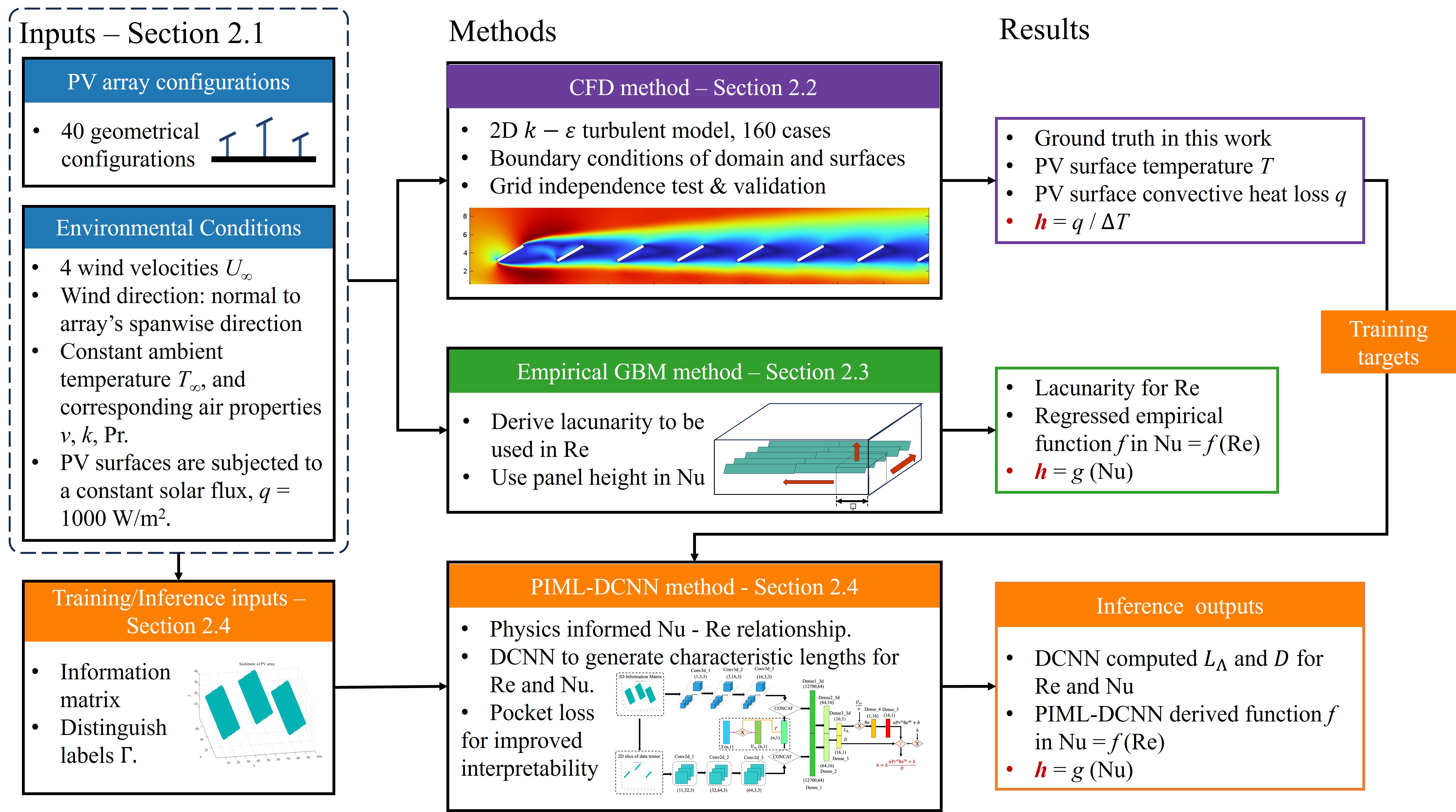}
  \caption{Methodology flowchart.}
  \label{fig:Schematics of Methodology}
\end{figure}

In this study, three methods are used to estimate heat transfer in PV arrays: an empirical method~\cite{smithview}, a validated CFD model, and the proposed PIML-DCNN. The results of the CFD model serve as the ground truth for training and testing the PIML-DCNN model, and as the benchmark for evaluating the accuracy of the empirical method. For all three methods, the inputs are the configurations and environment conditions of the PV array, including panel heights, wind velocities, air properties, and subjected solar flux. The outputs are heat transfer coefficients, $h$. The methodology flow chart is presented in Fig.~\ref{fig:Schematics of Methodology}.

\subsection{Problem description}
\begin{figure}[!htbp]
  \centering
  \includegraphics[width=0.9\textwidth]{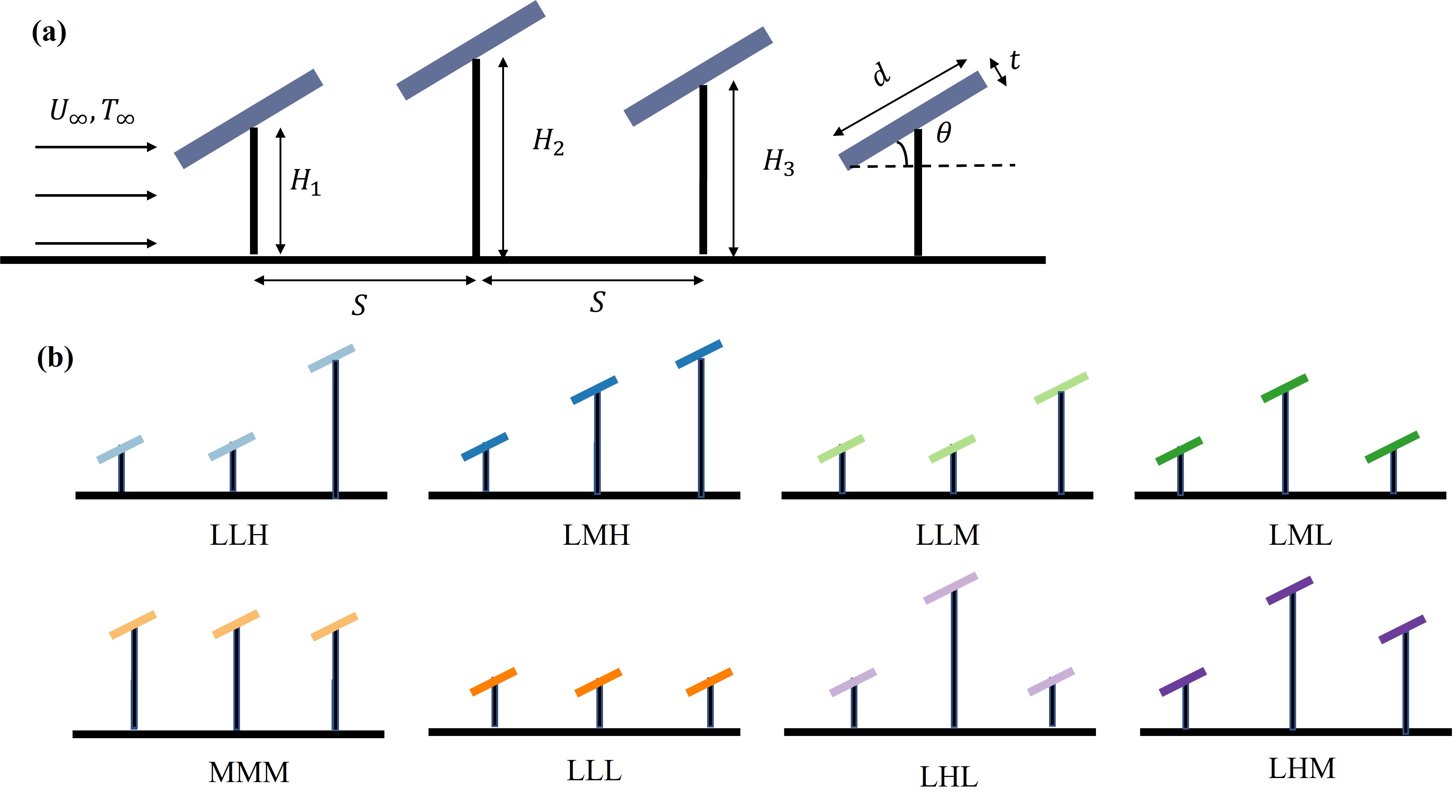}
  \caption{(a). Schematic of configurations settings. (b). Investigated height arrangement.}
  \label{fig:heights setting}
\end{figure}

Under the conditions of a constant ambient temperature $T_\infty$, a constant solar flux incident on PV panels, and a given wind speed $U_\infty$, the configuration of PV array -- especially the row spacing $S$ and the height $H$ of each row -- play a pivotal role in determining the turbulent flow and convective heat transfer performance~\cite{smithview}. 
In this study, the arrays are categorized into two different height configurations: the same-height array, where all rows have the same height, and the staggered-height array, which consists of rows with varying heights, as shown in Fig.~\ref{fig:heights setting}. To analyze the convective heat transfer rate under various conditions, the height arrangement, row spacing $S$, and wind speed $U_\infty$ are selected as manipulated variables. The variations in the spanwise direction is therefore assumed to be negligible. A total of 160 cases are analyzed based on the configurations outlined in Fig.~\ref{fig:heights setting}, with each case being a unique combination of height arrangement, $S$, and $U_\infty$. Other configuration parameters are determined in accordance with the dimensions reported in \cite{smithview}. A constant solar irradiance of 562.5 W/m$^2$ is applied to the upper surface of all PV panels~\cite{glick2020infinite}, with 112.5 W/m$^2$ being converted into electricity and the remaining 450 W/m$^2$ being dissipated through convection. The 450 W/m$^2$ of absorbed solar energy is equivalent to an internal heat source with an intensity of 1286 W/m$^3$, considering the dimensions of the considered PV panels. 

\subsection{The CFD method}\label{Subsection:Numerical Method}

The CFD method is used to compute both velocity and temperature fields by solving the momentum and energy equations using COMSOL Multiphysics. To account for turbulence effects on heat transfer rather than its dynamics details, $k-\varepsilon$ turbulence model is adopted. Boundary conditions for the CFD simulations are presented in Table\,\ref{Table: boundary contidion}. The fluid is assumed to be incompressible and its pressure is set as $p_{\infty}$ = 1\,atm. Open Boundary condition indicates a normal stress $f_0$ equals to zero. The size of the computational domain, the ambient temperature, and the initial temperature of the PV module are adopted from~\cite{Stanislawski}. The average convective heat transfer coefficient of the array, $h$, is calculated based on the average values of each panel's convective heat transfer rate and surface temperature,
\begin{equation}
h = \frac{\bar{q_s}}{{\bar{T}_{\rm surf}} - T_{\infty}}
\label{equation:governing equation}
\end{equation}
where $\bar{q_s}$ and ${\bar{T}_{\rm surf}}$ are the average convective heat flux and temperature of all panel surfaces, respectively.

\begin{table}[!htbp]
\caption{Boundary conditions of CFD simulation. The computation domain is $L$ = 300\,m and $H$ = 53\,m.}
\label{Table: boundary contidion}
\begin{tabular}{llllll}
\toprule
         & $x$ = 0 (inlet)                                 & ${x}/{L}=1$ (outlet) & $z= 0$ (bottom) & ${z}/{H}= 1$ (top)   & PV body \\ \hline
Momentum & $U = U_{\infty}$      & Open Boundary  & $U$ = 0\,m/s         & Open Boundary  & --      \\ 
Thermal  & $T_{\infty}$ = 300.15\,K & $T_0$ = 300.15\,K  & Zero flux     & $T_0$ = 300.15\,K & $q$ = 1286\,W/m$^3$  \\ 
\bottomrule 
\end{tabular}
\end{table}

The validation of CFD results is presented in the Appendix. The validated CFD model is then used to generate a dataset comprising 160 cases, each with different model inputs and outputs. Out of these cases, 120 cases are utilized to train the PIML-DCNN, and the remaining 40 cases are reserved for testing the model.

\subsection{The empirical method}\label{Subsection:Empricial method}

To estimate the convective heat transfer of PV arrays with inclined PV panels, the following empirical correlation is proposed by~\citet{smithview},

\begin{equation}
\text{log}_{10}(\text{Nu}) = a\text{Re}^m \text{Pr}^n + b
\label{equation:Sarah's equation}
\end{equation}
where $a$, $b$, $m$, and $n$ are empirically fitted coefficients. Pr is the Prandtl number, and Nu is the Nusselt number, representing the ratio between heat convection and conduction, 
\begin{equation}
\text{Nu} = \frac{hD}{k}
\label{equation:Nu}
\end{equation}
where, $h$ is the convective heat transfer coefficient, and $k$ is the thermal conductivity of air. $D$ is the characteristic height of the array, which is represented by the canopy height of the panel at the 8$^{\text{th}}$ row, where the flow boundary is fully developed~\cite{smithview, Stanislawski}.
Re is the Reynolds number, representing the ratio between inertial and viscous forces,
\begin{equation}
\text{Re} = \frac{U_{\infty}L_{\rm c}}{v}
\label{equation:Re}
\end{equation}
where $U_{\infty}$ represents the wind velocity, $L_{\rm c}$ is the characteristic length, and $v$ is the kinematic viscosity of air. 

\begin{figure}[htbp]
\centering
\includegraphics[width=0.6\textwidth]{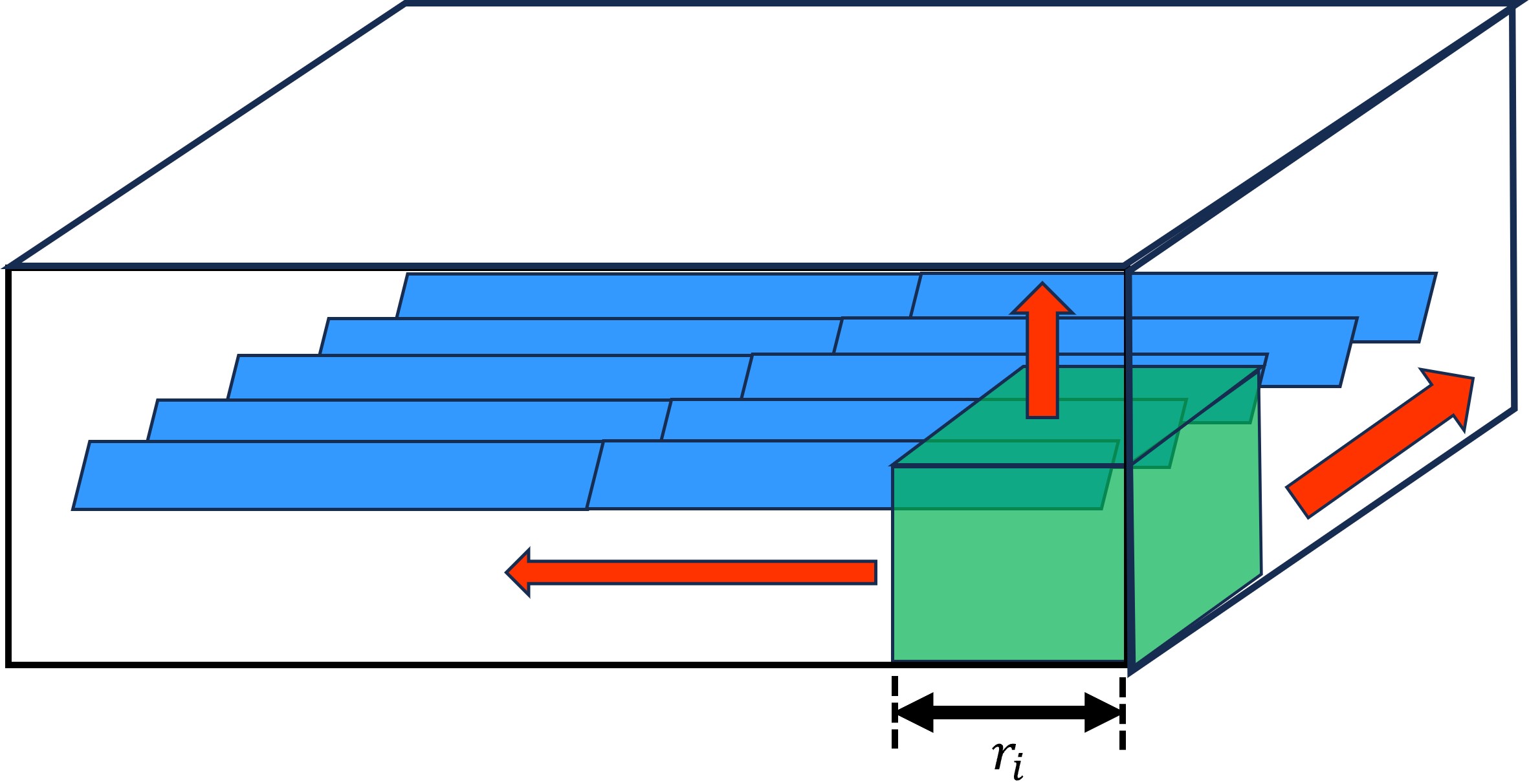}
\caption{Schematic illustration of the GBM for calculating lacunarity, where $r_i$ stands for the corresponding length of the box for quantifying $\Lambda$ at current step.}
\label{fig:GBM}
\end{figure}

To ensure the accuracy of the above empirical equation, it is essential to derive a characteristic length scale, $L_{\rm c}$, to represent the configurations of PV arrays, where PV panels are non-homogeneously distributed. Therefore, a method of deriving $L_{\rm c}$ based on the lacunarity concept is proposed by \citet{smithview}. Lacunarity is quantified by Gliding Box Method (GBM), where multiple boxes with different lengths are introduced to move through the whole PV array's space. GBM updates the PV configuration's occupation features within the box in each step and generates an overall lacunarity dimensionless number, $\Lambda$, from all the updates, as shown in Fig.~\ref{fig:GBM}~\cite{PhysRevE.53.5461,smithview}. Therefore, the characteristic length $L_{\rm c}$ can be represented as, 
\begin{equation}
\label{equation:lacunarity}
L_c = \frac{\sum (\Lambda \times R)}{N_R}, \quad R = [r_1, r_2, \ldots, r_N]
\end{equation}
where $R$ represents the set of box sizes used to calculate lacunarity, $\Lambda$ is the corresponding lacunarity of different boxes, $r_N$ is half of the smallest length of a PV panel, and $N_R$ is the number of  utilized box with different sizes~\cite{smithview}. 

\begin{figure}[!htbp]
\centering
    \includegraphics[scale=0.6]{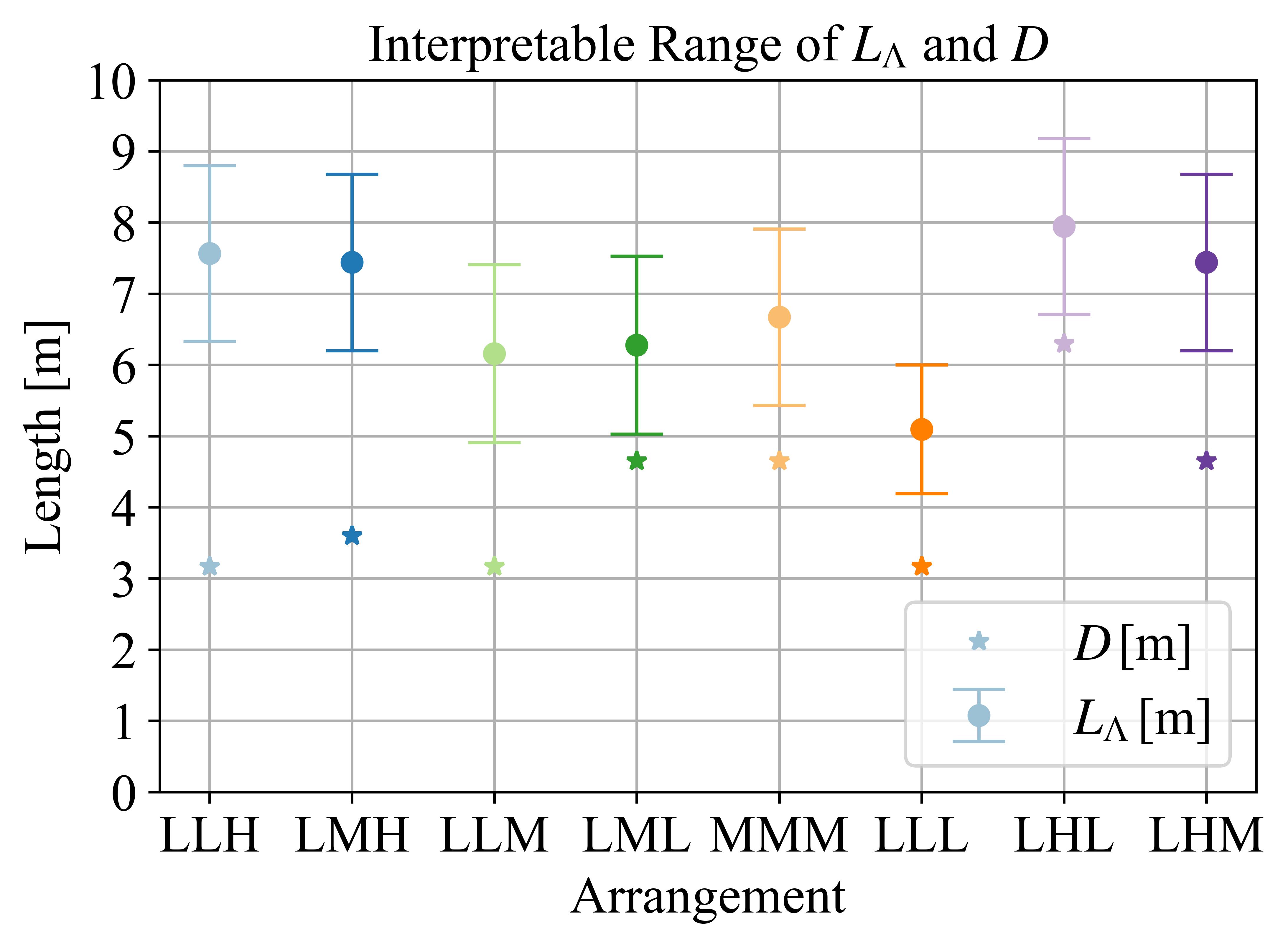}
\caption{The characteristic length $L_c$ and characteristic height $D$ derived from the empirical method for all investigated geometric configurations.}
\label{fig:interpretableRange}
\end{figure}

For the 160 cases we considered, the characteristic length and height obtained via the empirical method are illustrated in Fig.~\ref{fig:interpretableRange}. The results show that the characteristic length ranges from 4 to 10 m, while the characteristic height ranges from 3 to 7\,m, respectively.

Another widely used empirical correlation is~\cite{3d7d8616619744b2afe7aab5b1dcdc05, Elshamy2007EXPERIMENTALAN, BergamanHeatANDMass,smithview}, 
\begin{equation}
\text{Nu} = a\text{Re}^m \text{Pr}^n + b
\label{equation:Physical Knowledge}
\end{equation}
which will also function as the benchmark model against which the performance of the proposed PIML-DCNN model will be evaluated.

\subsection{The PIML-DCNN method}\label{Subsection:PIML-DCNN}

Although the CFD model can achieve satisfactory levels of accuracy, it requires substantial computational resources. Conversely, the empirical model provides efficient estimations but sacrifices accuracy. To substantially improve the accuracy of empirical model, it is crucial to effectively capture complex geometric configurations. Deep Convolutional Neural Networks (DCNNs) can serve as a robust alternative to GBM for this purpose, as both employ iterative updates within kernels to process geometric information. Here, we introduce the Physics-Informed Machine Learning-Deep Convolutional Neural Network (PIML-DCNN) as a solution that balances high accuracy with computational efficiency. 

In this model, a Deep Convolutional Neural Network (DCNN) is employed to extract values of $L_{\rm c}$ and $D$ from the complex configurations characteristic of both same-height and staggered-height arrangements. Specifically, $L_{\rm c}$ is determined using a 3D DCNN, while $D$ is obtained through a 2D DCNN. The values of $L_{\rm c}$ and $D$ thus derived are subsequently utilized to compute the Nusselt number (Nu) and Reynolds number (Re). Moreover, the dense layers of the deep learning (DL) model are designed to capture the empirical correlation between Nu and Re, as expressed in Eq.~(\ref{equation:Physical Knowledge}). This correlation is informed by physical principles and is instrumental in estimating the convective heat transfer coefficient, $h$.

\subsubsection{Model inputs and outputs}
\label{subsubsection: Data preparation}
The input data for the PIML-DCNN model encompass both environmental conditions and configurations of the photovoltaic (PV) array. The output generated by the model is the convective heat transfer coefficient, denoted as $h$. The learning objective for the model is to match this output with the corresponding $h$ values derived from the CFD model.

\begin{figure}[!htbp]
    \centering
    \includegraphics[scale=0.35]{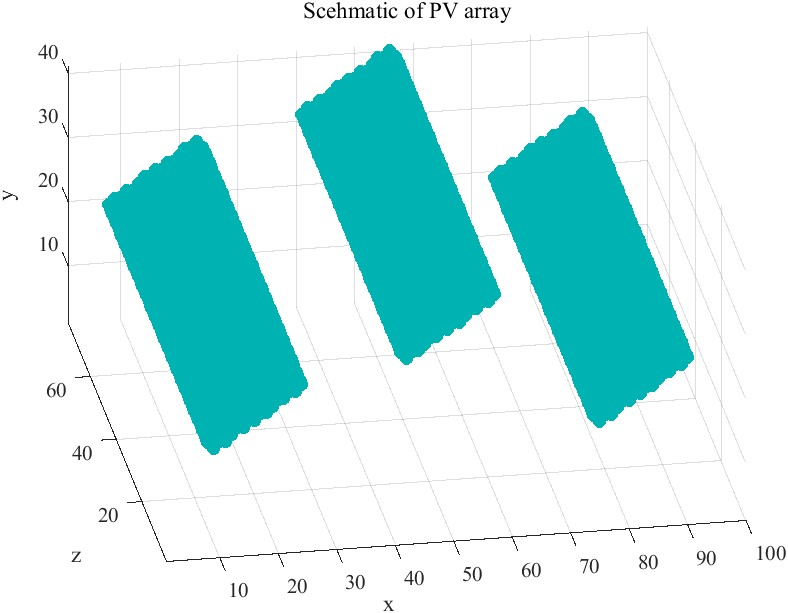}
    \caption{The information matrix of 3 rows of PV panels.}
    \label{fig:PV geometry}
\end{figure}

The information matrix is constructed with a resolution that is half of the minimum dimension among all input features, following the guidelines proposed in~\cite{smithview}. This matrix captures essential characteristics of PV arrays, such as the number of rows, panel dimensions, and spacing between rows. In this matrix, geometric shapes are represented by discrete points, with a `1' indicating the presence of a panel and a `0' denoting its absence. These points are then plotted in a 3D space to form the information matrix, which graphically represents the layout of the PV panels, as depicted in Fig.~\ref{fig:PV geometry}. It is important to note that, for our PIML-DCNN model, configurations of 10 rows are employed, as opposed to the 3-row configuration illustrated in Fig.~\ref{fig:PV geometry}.
The spanwise length of the panel within the information matrix is configured to be 2.0 m, aligning with the specifications referenced in \cite{Stanislawski}.

To consider the effects of wind velocity $U_{\infty}$ and row spacing $S$ in the PIML-DCNN model, a parameter named $\Gamma$ is defined,
\begin{equation}
\label{equation:gamma}
    \Gamma = S \times U_{\infty} 
\end{equation}
which is introduced to consider the turbulent field features and amplify row spacing effect while saving the effort from reconstructing high-resolution data with all detailed configuration features.

\subsubsection{PIML-DCNN model}
\begin{figure}[!htbp]
    \centering
    \includegraphics[scale=0.28]{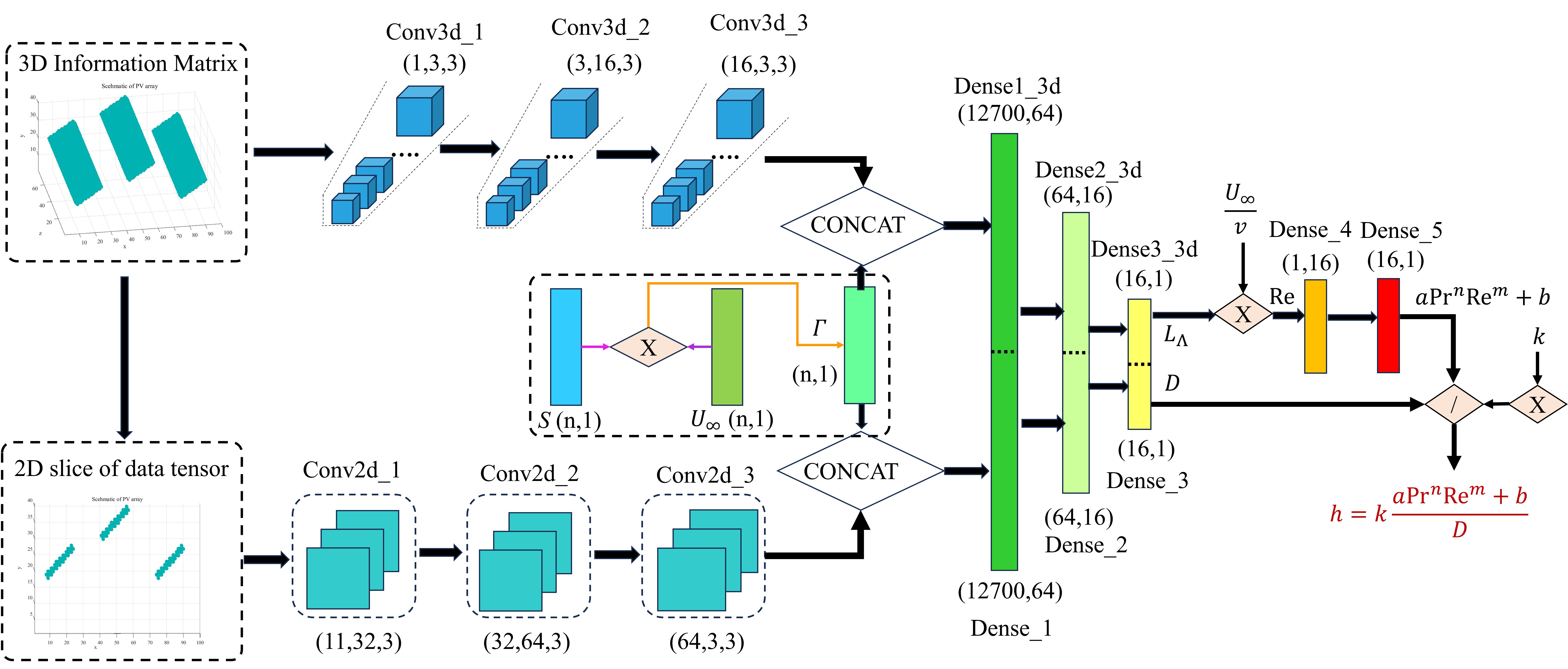}
    \caption{Architecture of the Developed PIML-DCNN Model. The tuple represents the size of each layer, with the final number indicating the kernel size for the DCNN layers.}
    \label{fig:PIML-DCNN layer}
\end{figure}

The inputs of the PIML-DCNN model are: 3D information matrix, 2D slice matrix as the projection on the wind field plane, $\Gamma$, $U_{\infty}/v$, and $k$. As shown in Fig.~ \ref{fig:PIML-DCNN layer}, the 3D information matrix is fed into a 3D DCNN structure and then go through feed forward layers, which is labeled as $\theta$, to generate $L_{\rm c}$,
\begin{equation}
    L_c = \theta_{\rm Dense1\_3d - Dense3\_3d} (\Phi_{\rm 3D} (\Psi_{\rm 3D}))
\label{equation:3D CNN}
\end{equation}
The 2D slice matrix goes through a similar process to output $D$,
\begin{equation}
    D = \theta_{\rm Dense1 - Dense3} (\Phi_{\rm 2D} (\Psi_{\rm 2D}))
\label{equation:2D CNN}
\end{equation}

During the feed forward process of 3D DCNNs and 2D DCNNs, after convolution, the $\Gamma$ parameter will be combined with both 3D and 2D convolution layers outputs to form the input of forward layers to generate $L_c$ and $D$. Then $L_{\rm c}$ will combine with $U_{\infty}/v$ to get Re, which will then goes through layers Dense4 and Dense5 to approximate the empirical correlation Eq.~(\ref{equation:Physical Knowledge}) ~\cite{SVOZIL199743},
\begin{equation}
    \text{Nu} = a\text{Re}^m\text{Pr}^n+b = \left(\prod_{i = 1}^2 w_{3-i}^\text{T}\right)\text{Re}+b
    \label{equation:approximator}
\end{equation}
Detailed proof of this approach is presented in the Appendix. 
Finally, $h$ will be calculated using Nu, $D$ and $k$.
 
Drawing inspiration from the U-Net architecture, which has shown efficacy in processing low-resolution images as demonstrated in~\cite{ronneberger2015unet}, our approach employs a similar strategy for handling information matrix data, which can be considered a form of low-resolution imagery. In each convolutional layer, we introduce a subsequent max pooling layer to effectively downsample the data. Additionally, we implement an Early Dropout technique to mitigate underfitting when working with datasets of limited size during training~\cite{hinton2012improving}. The dropout layer is strategically active only during the initial 20\% of the training epochs and is then deactivated for the remainder of the training period.

\subsubsection{Pocket loss function}

Although the current PIML-DCNN model can accurately approximate $h$ in 160 cases, the resulting values for $L_{\rm c}$ range from -0.02 m to -0.01 m, and for $D$, from 2.26 m to 2.40 m. These ranges exceed those with physical relevance as detailed in Section~\ref{Subsection:Empricial method}. The non-physical nature of $L_{\rm c}$ and $D$ indicates the model's limited interpretability, particularly concerning the Nusselt number. Enhancing the interpretability of PIML may involve customizing loss functions, such as incorporating physical laws directly into the loss function, as suggested in ~\cite{ZHOU2023108835,REN2023128472}, along with other innovative approaches like the one proposed in ~\cite{RAO2020207}. Consequently, we introduce a novel loss function, termed Pocket Loss, to address these issues,

\begin{equation}
    \xi_i = \frac{{\bar \chi_i}^3}{{C_{\rm up}}^3} + \frac{1}{e^{\bar \chi_i - C_{\rm low}}}
\end{equation}

where $\xi_i$ represents the Pocket Loss for the physical parameter requiring improved interpretability; $\bar \chi_i$ denotes the average value of $\chi_i$. The constants $C_{\rm up}$ and $C_{\rm low}$ are the empirical or domain knowledge-based upper and lower boundary values for $\chi_i$, respectively. We define Pocket Loss in this manner to maintain the model's overall convexity during training — particularly important when generating negative outputs — and to robustly encourage exploration around the defined boundary values. This is achieved because the gradients near the boundaries are significantly lower than those further away, guiding the model to stay within realistic limits.

\begin{figure}[!htbp]
    \centering
    \includegraphics[scale=0.5]{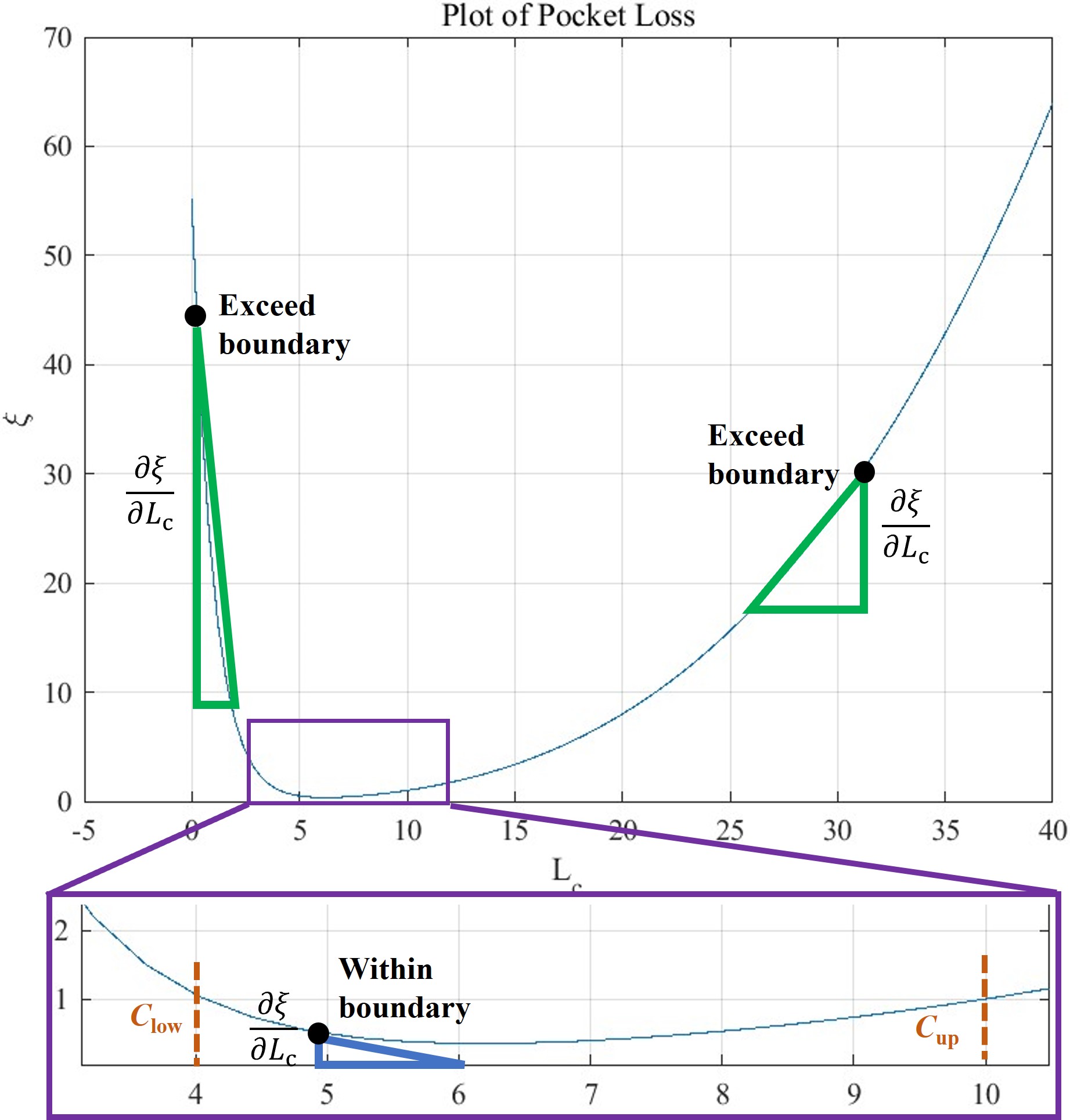}
    \caption{Illustration of Pocket Loss for $L_{\rm c}$ with Gradient Representation. The plot displays the loss values as dots, with gradients indicated by triangles: green for values exceeding the boundary and blue for those within. Dashed lines highlight the lower and upper boundary values.}
    \label{fig:Pocket Loss}
\end{figure}

Taking $L_{\rm c}$ as an example, its range is approximately between 4 m and 10 m as depicted in Fig.~\ref{fig:interpretableRange}. We set the constraint boundaries with $C_{\rm up} = 10$ m and $C_{\rm low} = 4$ m to ensure $L_{\rm c}$ remains within this specified range. As illustrated in Fig.~\ref{fig:Pocket Loss}, during training, values that exceed the boundaries are subjected to a higher magnitude and steeper gradients of loss compared to values that stay within the boundaries, where the loss remains below 1.0. Consequently, as the deep learning model aims to minimize the loss value and its gradient, the upper and lower boundaries act as a `pocket', containing the changes in $L_{\rm c}$ and its gradient. This pocket guides the model towards the defined range, resulting in relatively low values of $\xi$ and $\partial \xi/{\partial L_{\rm c}}$. Additionally, Pocket Loss remains differentiable within the boundary region, which circumvents issues related to non-convexity and local minima, thus ensuring a smooth and robust training process.

The boundary value settings for $D$ are also based on the information presented in Fig.~\ref{fig:interpretableRange}, with $C_{\rm up}$ set to 7 m and $C_{\rm low}$ set to 3 m. When interpretability is required for multiple physical parameters, Pocket Loss is redefined as,

\begin{equation}
\xi = \sqrt{\sum_{i = 1}^n \left(\frac{{\bar \chi_i}^3}{{C_{{\rm up},i}}^3} + \frac{1}{e^{\bar \chi_i - C_{{\rm low},i}}}\right)^2}
\end{equation}

This equation directs the adaptation of all parameters during training by minimizing the projected distance within an n-dimensional geometric space—where $n$ represents the number of physical parameters requiring interpretability. The refined loss function for training the model is subsequently defined as follows:

\begin{equation}
\text{Loss} = \text{RMSE} + \alpha \xi
\end{equation}
where RMSE represents the Root Mean Square Error, which quantifies the discrepancy between the model's predictions and the actual target values. The coefficient $\alpha$ is introduced to calibrate the gradient balance between $\xi$ (the interpretability term) and RMSE during training. For our experiments, we have determined the value of $\alpha$ to be 0.1.

To enhance the smoothness and stability of the training process, we employ a total of 300 epochs. During the initial 50 epochs, we use a learning rate of $10^{-3}$ in conjunction with the Adam optimizer and early dropout techniques to swiftly guide the model towards the predefined boundaries. Subsequently, the learning rate is decreased to 
$10^{-4}$ for the second phase of 150 epochs and to 
$10^{-5}$ for the final 100 epochs. These adjustments, coupled with the continuation of early dropout techniques, are aimed at fine-tuning the model to achieve optimal results within or in close proximity to the established boundaries.

\section{Results and Discussion}\label{Sec:ResultsandDiscussion}

\subsection{Performance of the empirical method}

To assess the performance of both the empirical method and the PIML-DCNN method when taking the CFD results as ground truth, the coefficient of determination ($R^2$) and the maximum relative error ($\varepsilon$) are utilized as metrics, which is defined as:

\begin{equation}
\label{equation:maxRelativeError}
\varepsilon = \left( \frac{\text{MAE}(\hat y, y)}{\min(\bar y, \bar{\hat y})} \right) \times 100\%
\end{equation}

In this context, $\hat y$ denotes the predicted value, $y$ represents the value obtained from the CFD simulation (ground truth), $\bar y$ and $\bar{\hat y}$ are the average values of $y$ and $\hat y$ respectively, and MAE is the Mean Absolute Error between the predictions and the target values.

Upon applying the empirical model, as detailed in Eqs.~(\ref{equation:Sarah's equation}) and (\ref{equation:Physical Knowledge}), to our CFD dataset, the highest model accuracy was achieved with the parameters $a= 0.09$, $b= 1.91$, $m = 0.2$, and $n = 0.0833$ for Eq.~(\ref{equation:Sarah's equation}), and $a= 0.6093$, $b= 1.0597$, $m = 0.6336$, and $n = 1.3322$ for Eq.~(\ref{equation:Physical Knowledge}). Despite this optimization, the overall accuracy remains unsatisfactory, with an $R^2$ of 0.61 and a $\varepsilon$ of 23\%. This level of accuracy suggests that the characteristic lengths used in the empirical model may not adequately capture the complexity of PV array configurations.

\subsection{Performance of the PIML-DCNN model}\label{subsection:Model performance}

\begin{figure}[!htbp]
  \centering
  \begin{subfigure}[b]{0.48\textwidth}
    \includegraphics[width=\textwidth]{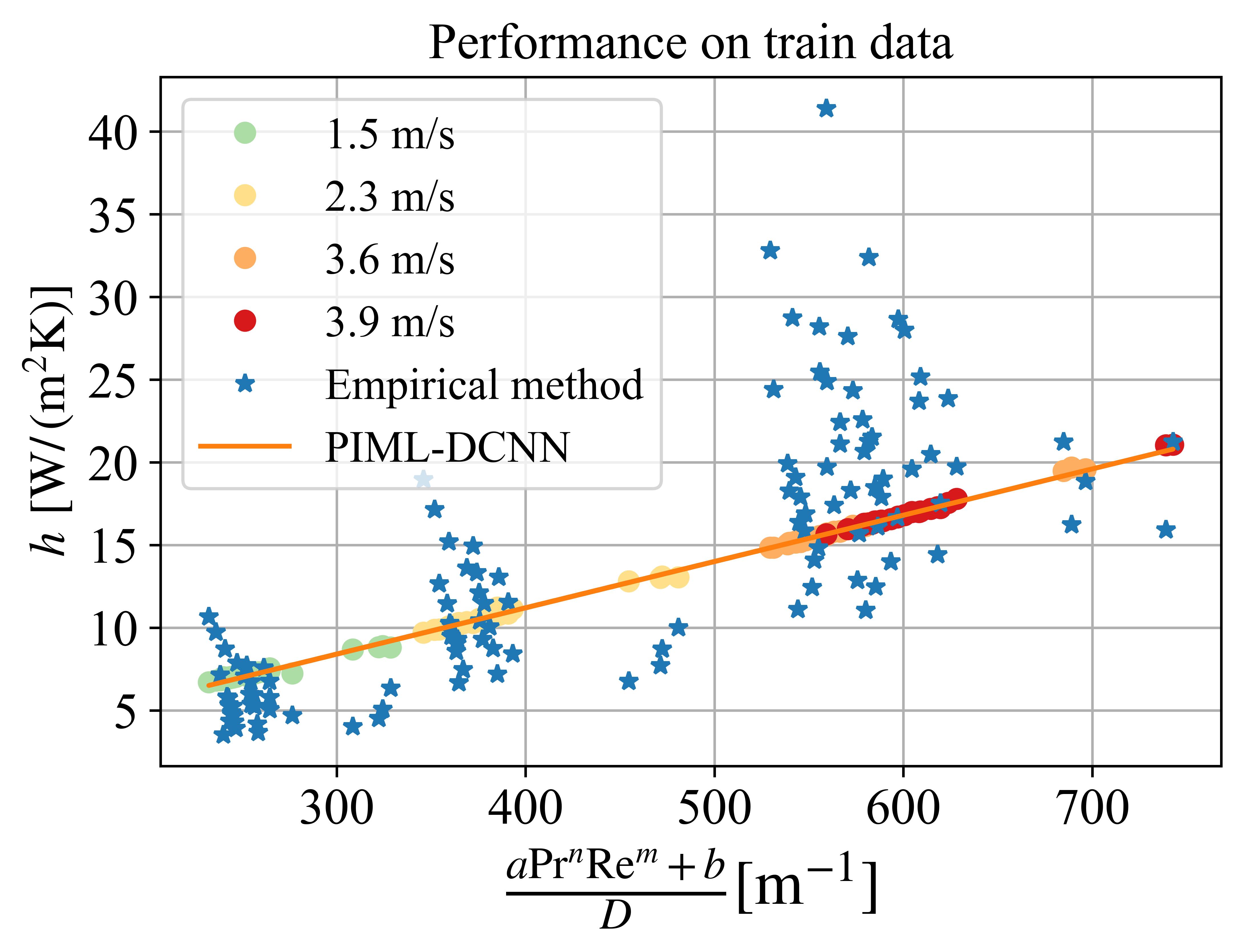}
  \end{subfigure}
  \hfill
  \begin{subfigure}[b]{0.48\textwidth}
    \includegraphics[width=\textwidth]{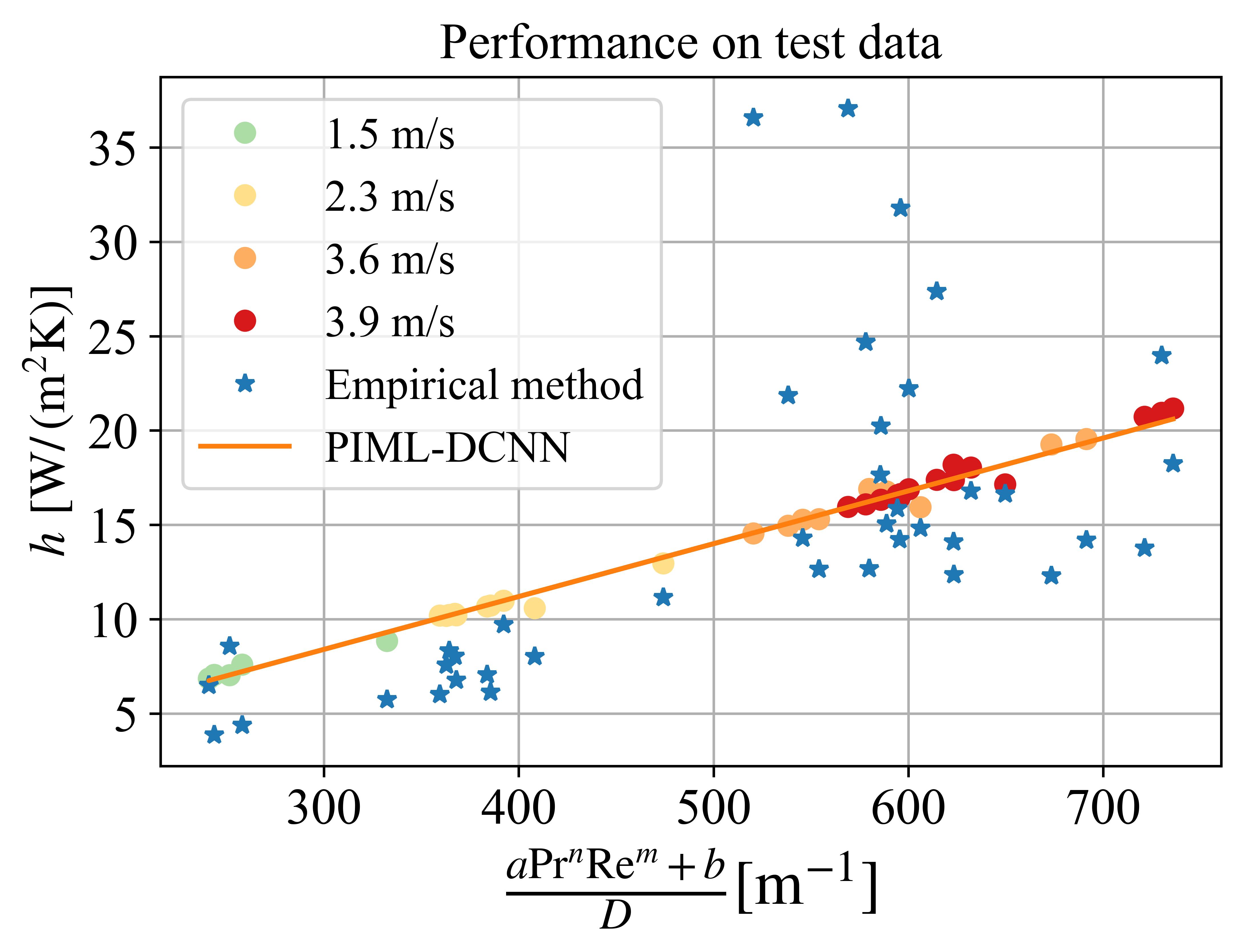}
  \end{subfigure}
  \caption{Performance of the proposed PIML-DCNN model with Pocket Loss function. Colored dots represent the CFD results under different wind speeds. Stars represent the results from the empirical method. The line represents PIML-DCNN predictions.}
  \label{fig:PIML-DCNN Performance}
\end{figure}

As illustrated in Fig.~\ref{fig:PIML-DCNN Performance} and detailed in Table~\ref{Table: Performance results}, the PIML-DCNN models exhibit outstanding performance on the testing dataset, achieving relative errors below 2\% and an $R^2$ value of 0.99. This indicates that the proposed PIML-DCNN models are significantly more accurate compared to the empirical model.

Furthermore, Table~\ref{Table: Performance results} demonstrates that the introduction of the Pocket Loss function can improve both the interpretability and the physical plausibility of the PIML-DCNN model. By incorporating Pocket Loss, the model ensures that the characteristic height $D$ is constrained to a physical range of 5.68 m to 6.86 m, while also keeping the characteristic length $L_{\rm c}$ within a physical range of 3.85 m to 4.87 m. Additionally, the Reynolds numbers predicted by the model incorporating Pocket Loss fall within a range of $3.81 \times 10^5$ to $1.17 \times 10^6$, which is in close agreement with the valid range established by \citet{smithview}.

In addition, the proposed PIML-DCNN model also requires less computational time than CFD approaches. For the CFD model used in this study, a single CFD case employing the $k-\varepsilon$ turbulence model requires at least 12 minutes to yield reasonable results on a high-performance AMD Ryzen Threadripper 3970X 32-Core Processor. In contrast, our trained PIML-DCNN model is capable of estimating the heat transfer performance of various configurations almost instantaneously.

\begin{table}[!htbp]
\caption{Prediction accuracy for the convective heat transfer coefficient ($h$) across various methods.}
\label{Table: Performance results}
\begin{tabular}{ccccc}
\hline
& $\varepsilon$ (\%) & $R^2$ &$L_{\rm c}$ & $D$    \\ \hline
Empirical Method of Eq.(\ref{equation:Sarah's equation}) & 23\% & 0.49 &[4.19\,m, 9.18\,m] & [3.17\,m, 6.21\,m]  \\
Empirical Method of Eq.(\ref{equation:Physical Knowledge}) & 23\% & 0.61  & [4.19\,m, 9.18\,m] & [3.17\,m, 6.21\,m]  \\ 
DCNN-PIML (without Pocket Loss)& 0.9\% & 0.99 & [-0.02\,m, -0.01\,m] & [2.36\,m, 2.40\,m]  \\ 
DCNN-PIML (with Pocket Loss) & 1.3\% & 0.99 & [3.85\,m, 4.87\,m] & [5.68\,m, 6.86\,m]  \\ \hline
\end{tabular}
\end{table}

\begin{figure}[!htbp]
    \begin{subfigure}[b]{0.5\textwidth}
    \includegraphics[scale = 0.55]{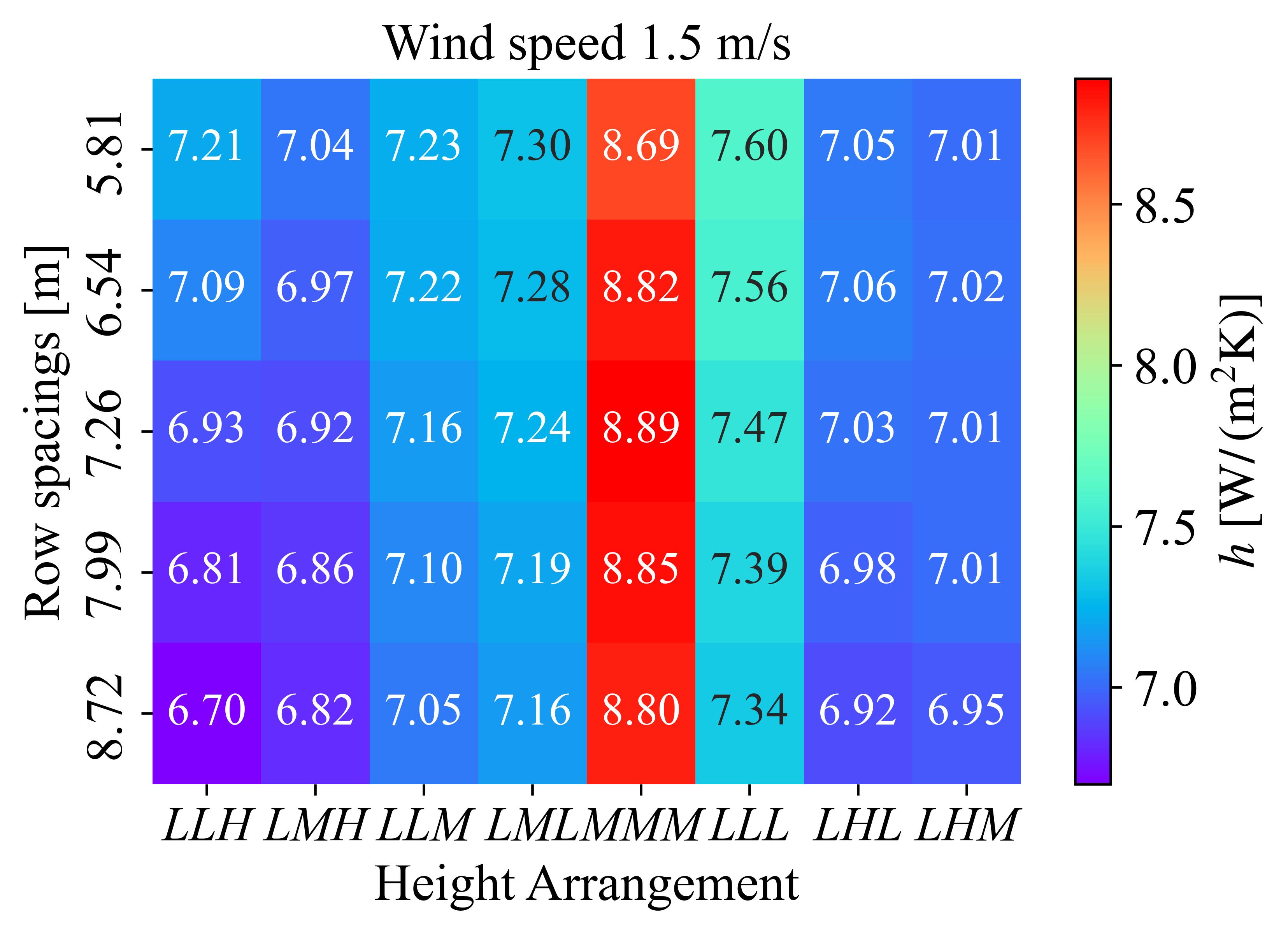}
  \end{subfigure}
  \begin{subfigure}[b]{0.5\textwidth}
    \includegraphics[scale=0.55]{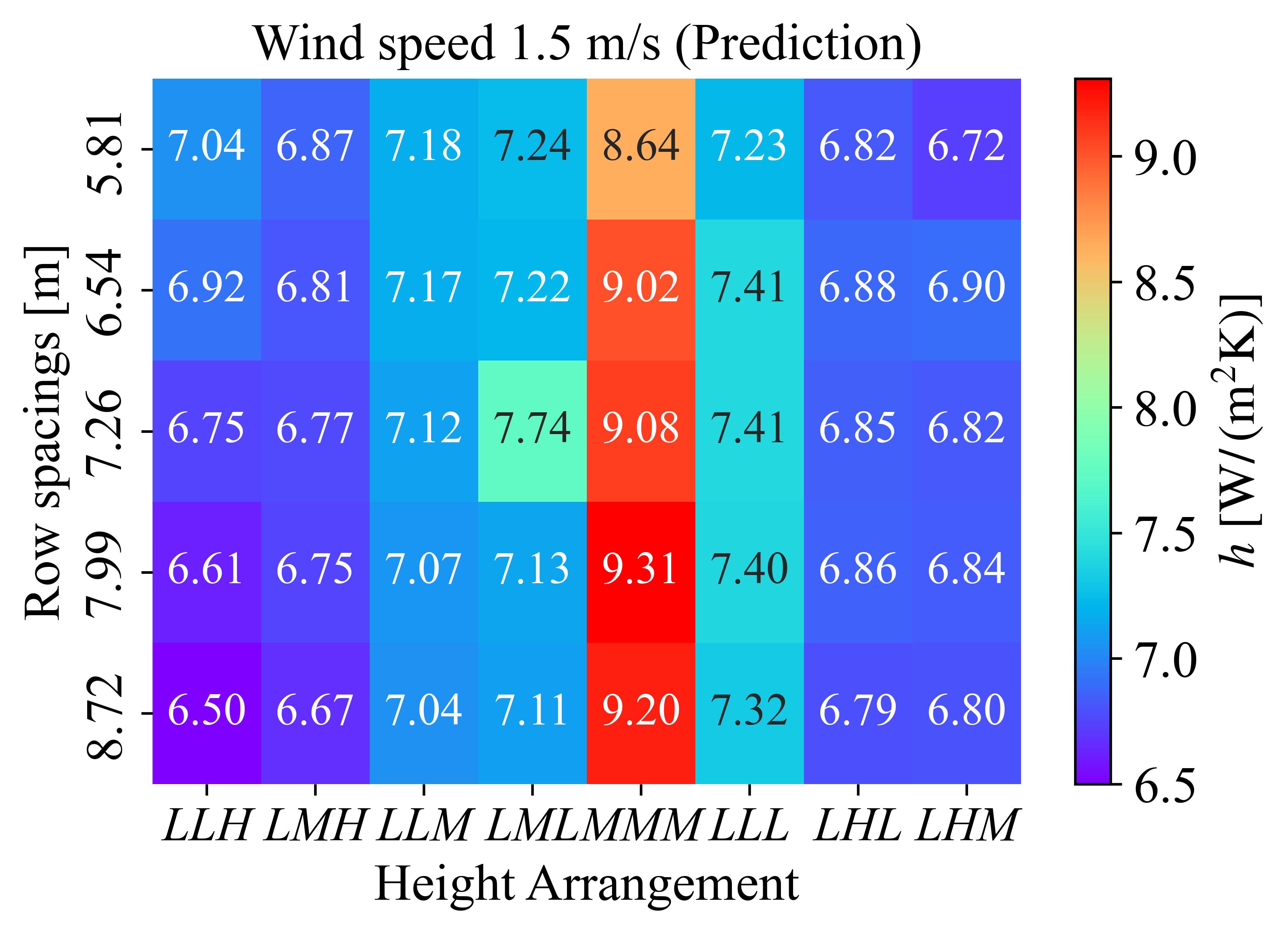}
  \end{subfigure}
      \begin{subfigure}[b]{0.5\textwidth}
    \includegraphics[scale = 0.55]{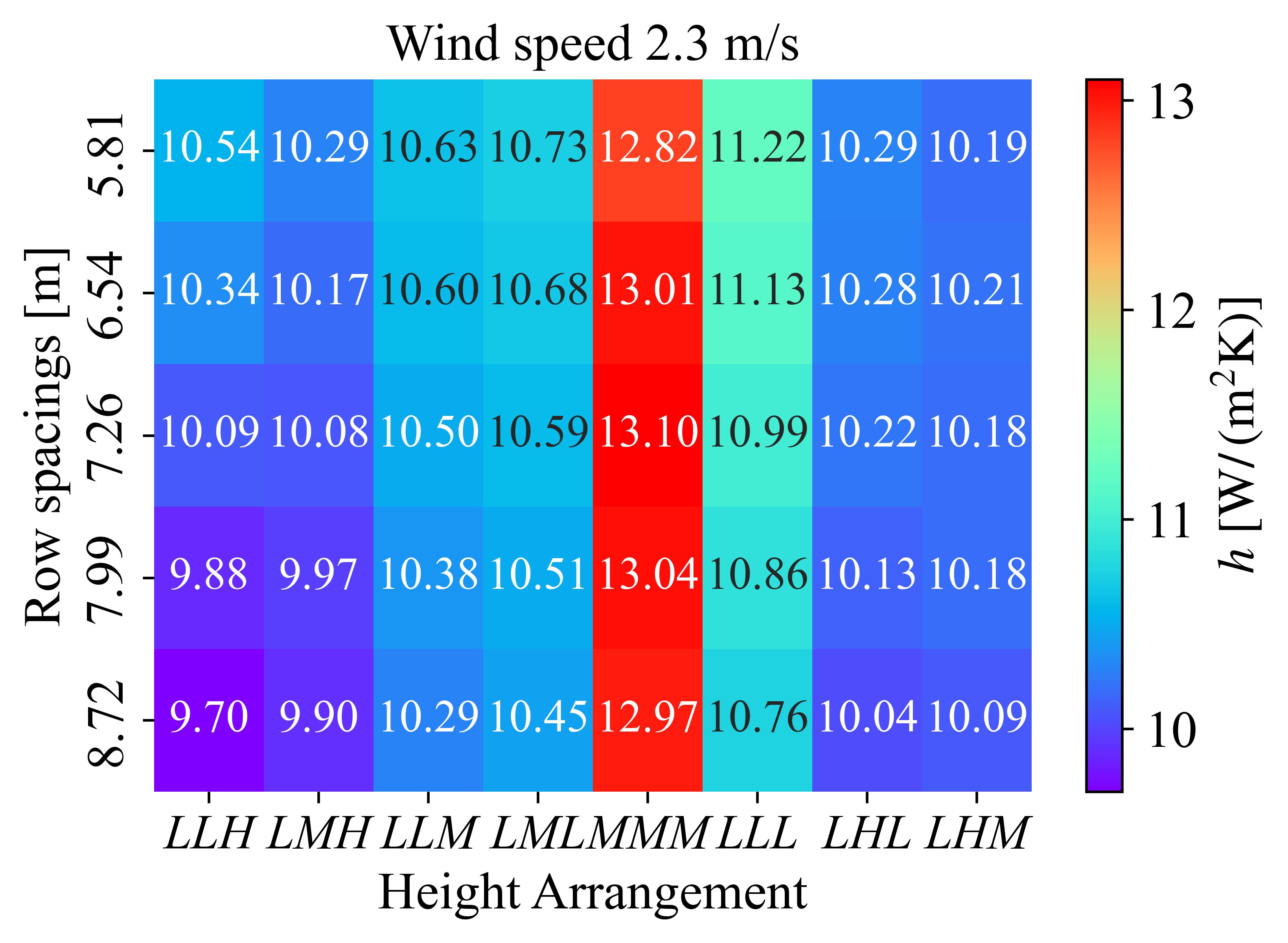}
  \end{subfigure}
  \begin{subfigure}[b]{0.5\textwidth}
    \includegraphics[scale=0.55]{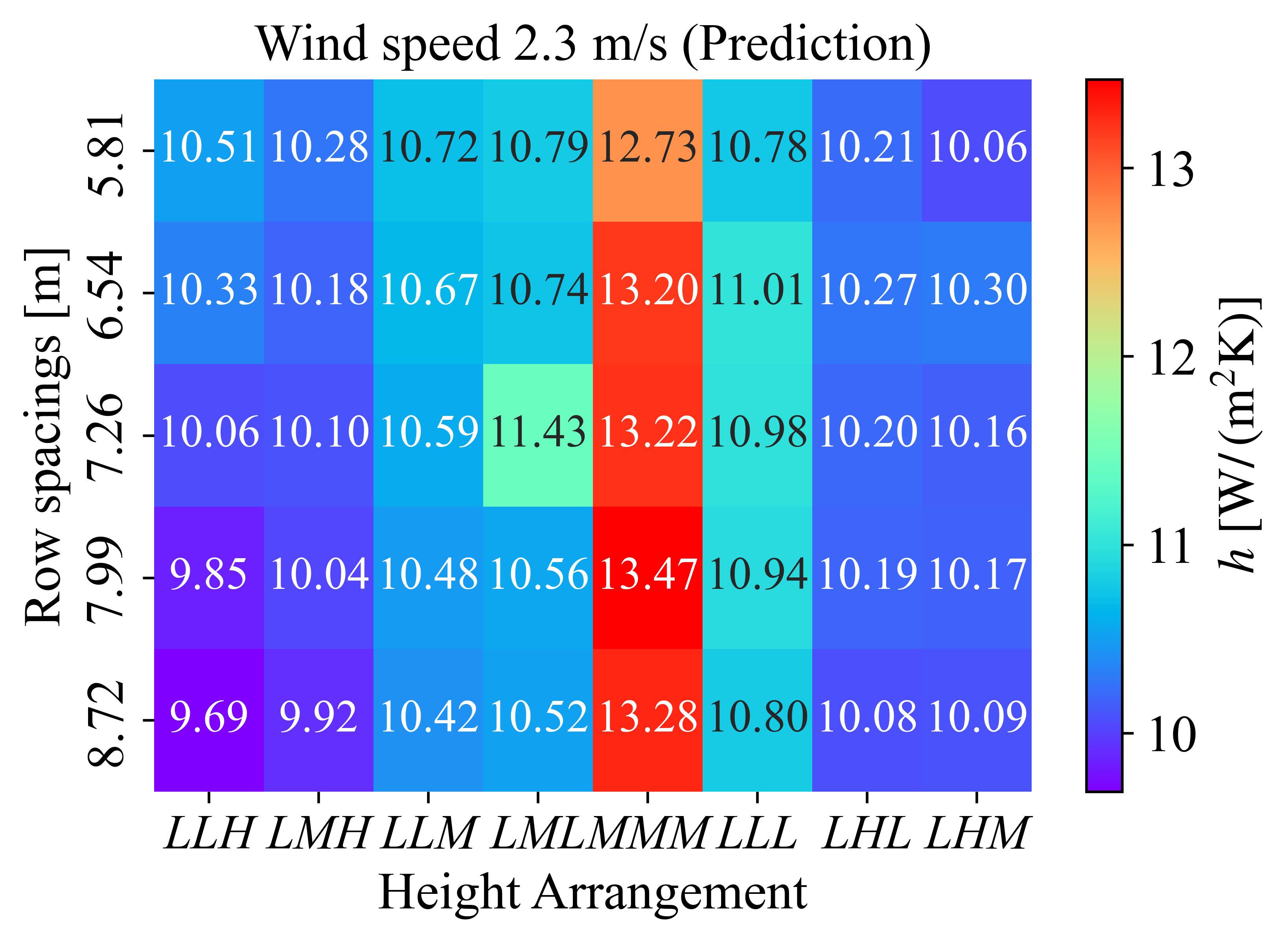}
  \end{subfigure}
    \begin{subfigure}[b]{0.5\textwidth}
    \includegraphics[scale=0.55]{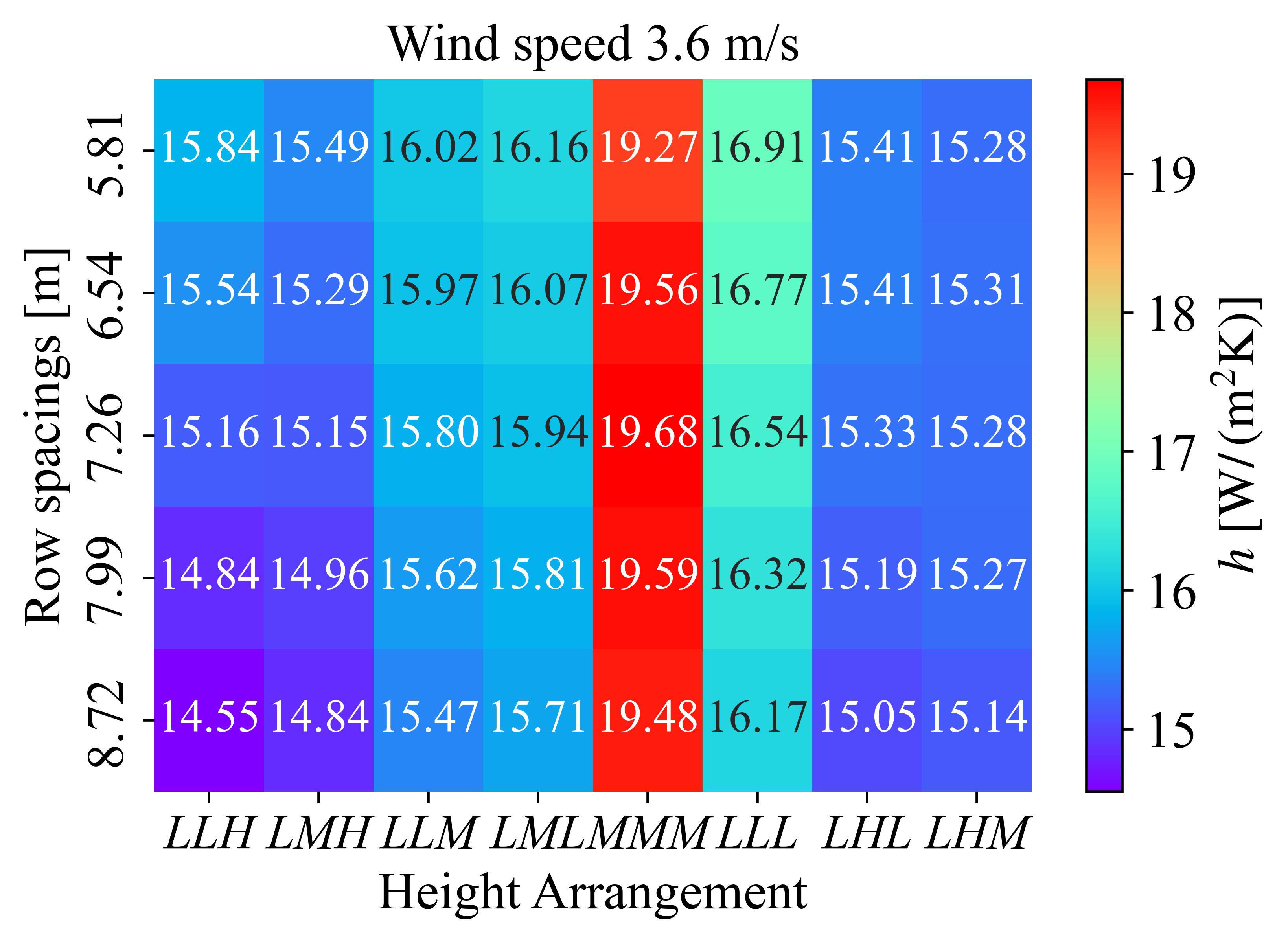}
  \end{subfigure}
    \begin{subfigure}[b]{0.5\textwidth}
    \includegraphics[scale=0.55]{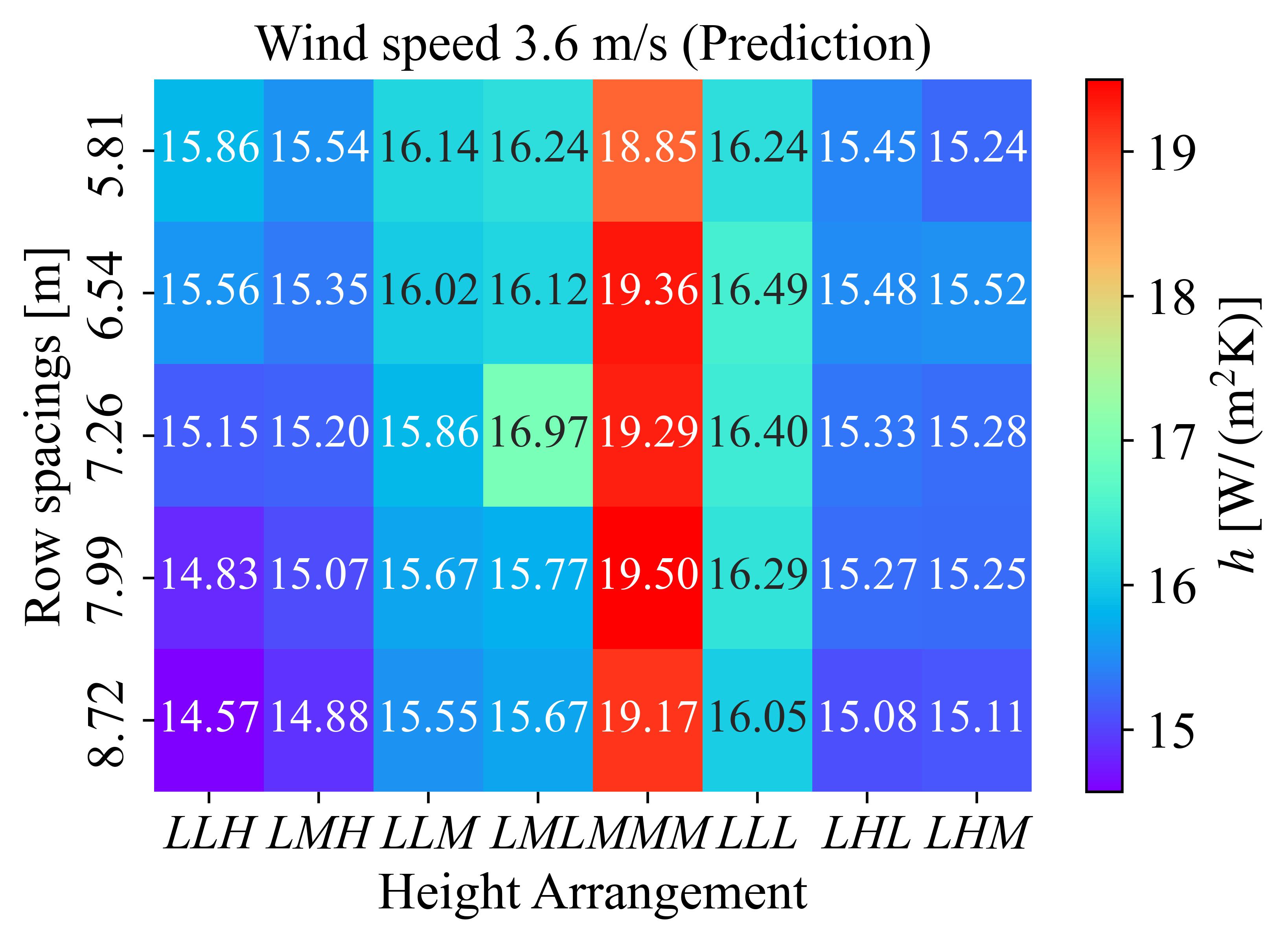}
  \end{subfigure}
    \begin{subfigure}[b]{0.5\textwidth}
    \includegraphics[scale=0.55]{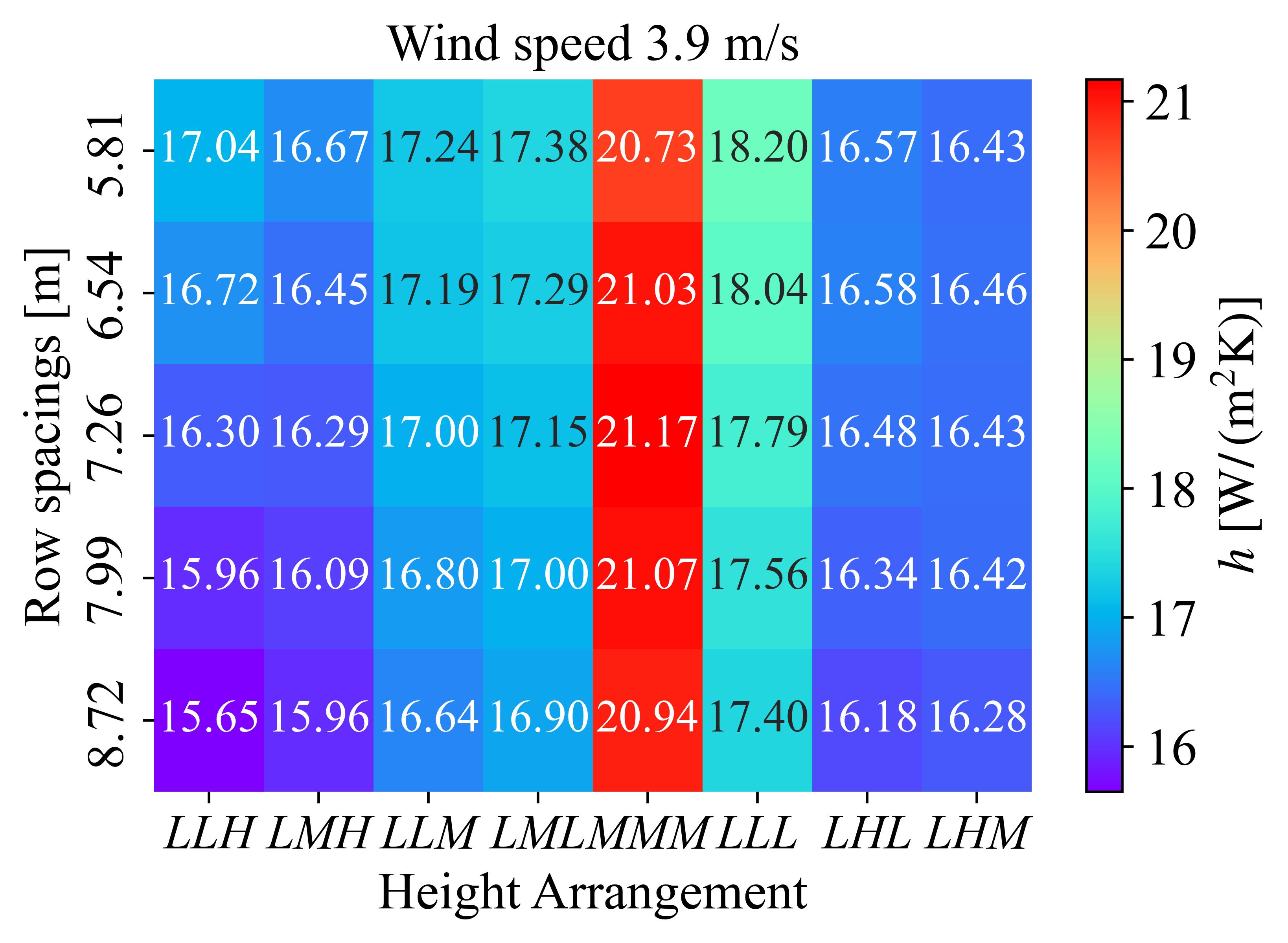}
  \end{subfigure}
    \begin{subfigure}[b]{0.5\textwidth}
    \includegraphics[scale=0.55]{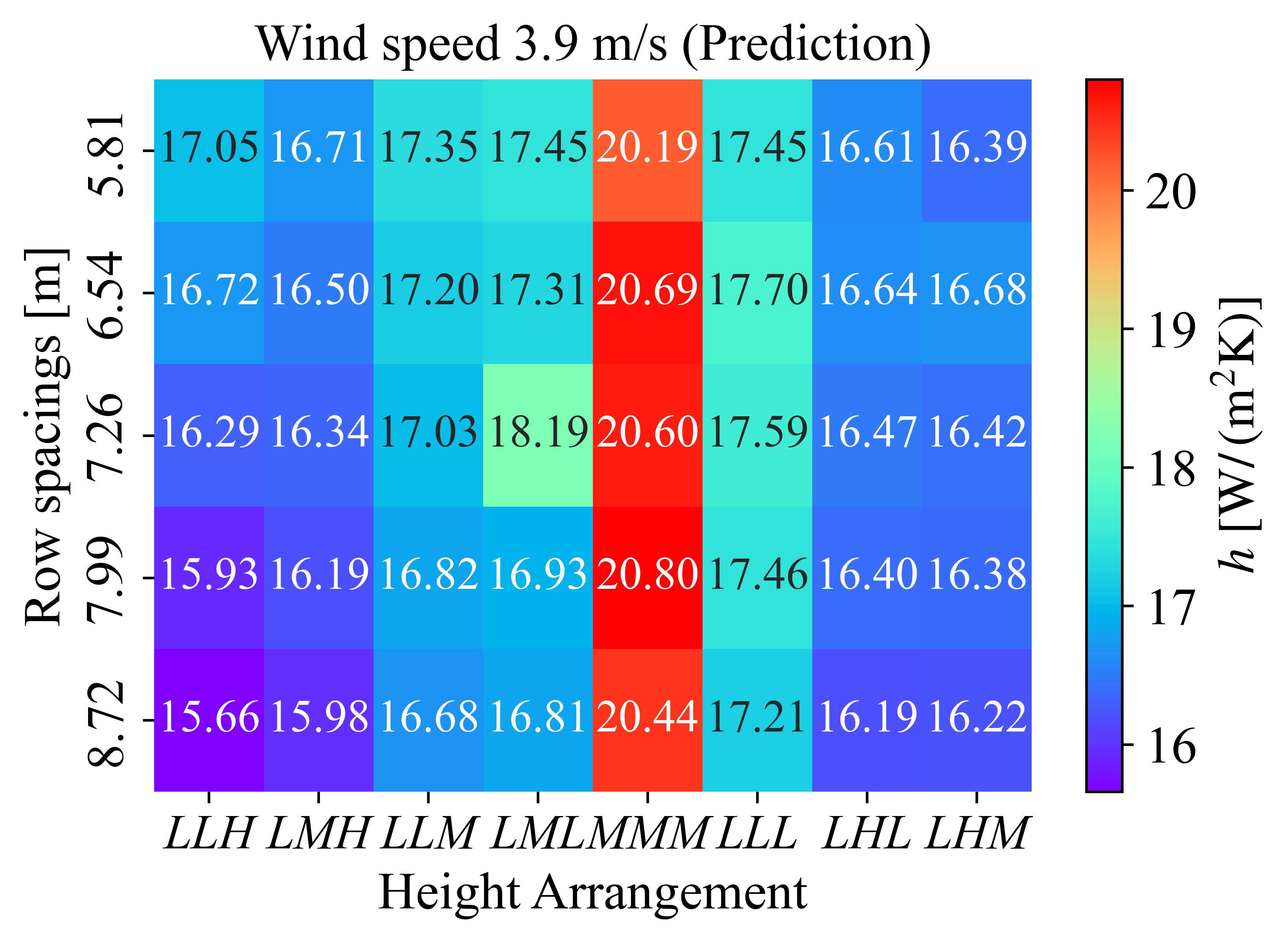}
  \end{subfigure}
  \caption{ Estimated $h$ from CFD method (left column) and PIML-DCNN method (right column) under different conditions. }
  \label{fig:heatmaps}
\end{figure}

Figure~\ref{fig:heatmaps} presents a visual comparison between the convective heat transfer coefficient ($h$) estimations from both CFD simulations and the PIML-DCNN model. The heatmaps indicating the average $h$ across the array from CFD results and model predictions show a high level of concordance, validating the PIML-DCNN model's accuracy. Additionally, Fig.~\ref{fig:heatmaps} reveals that staggered-height configurations encounter more significant heat transfer challenges than uniform-height scenarios, with the optimal arrangement being when all panels are at the same height of 3.3 m. However, the subpar performance of staggered-height configurations contrasts with their observed beneficial impact on regional flow optimization, as seen in Fig.~\ref{fig:CFDs}. To explain this inconsistency, a more in-depth analysis of the CFD results is undertaken in the subsequent section.


\subsection{Comparison between panel configurations}

\begin{figure}[!htbp]
    \centering
    \includegraphics[width=0.6\linewidth]{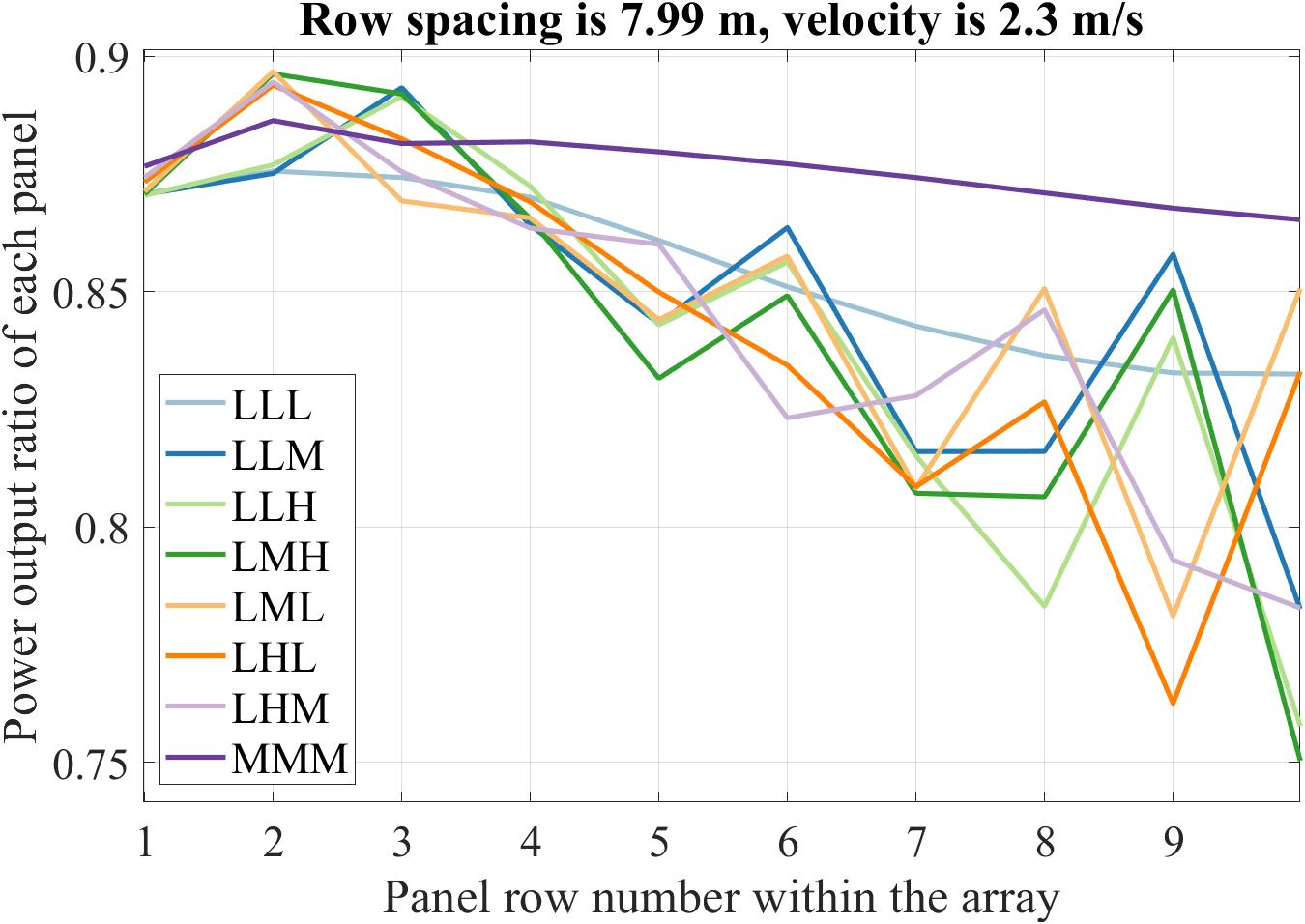}
    \caption{Relative power output for each panel in arrays with diverse configurations: all cases maintain identical row spacing of 7.99 m and a wind velocity of 2.3 m/s. Other scenarios exhibiting varied row spacings and wind velocities display comparable trends.}
    \label{fig:power analysis}
\end{figure}

The efficiency and power output of PV panels are significantly influenced by their temperature. To assess this impact, the power output ratio of panels with various configurations, as compared to the power output under standard conditions, is presented in Fig.~\ref{fig:power analysis}. The power output ratio is calculated using the panel temperatures determined by the CFD model ~\cite{Vaillon2018PathwaysFM},
\begin{equation}
r_{\text{PO}} = \frac{P}{P_{\rm STC}} = 1 + \beta_p(T_{\rm mod} - T_{\rm STC})
\end{equation}
where $r_{\text{PO}}$ is the power output ratio, $P$ is the power output, $P_{\rm STC}$ is the estimated power generation at the standard test condition (STC), $T_{\rm mod}$ is the panel temperature from CFD results, $T_{\rm STC}=25\,^{\circ}$C, and $\beta_p$ is the temperature coefficient of the panel, which is taken to be $-0.45\,\%/\text{K}$ \cite{Vaillon2018PathwaysFM}. Figure~\ref{fig:power analysis} illustrates that, in arrays comprising ten rows with staggered-height configurations, the first three rows demonstrate enhanced power generation compared to their counterparts in the uniform-height scenarios. Nevertheless, from the forth to the tenth rows, there is a noticeable decrease in heat transfer efficiency. This decline in performance for the panels situated downstream is more significant in the staggered-height configurations than in those with uniform heights.

\section{Conclusion}
\label{sec:conclusion}
The efficient transfer of heat through convection plays a vital role in the functionality of photovoltaic (PV) systems, given that panel temperature can affect PV power output. The layout of PV panels can markedly influence the convective heat dissipation by modifying the air flow. Traditional approaches to assess the impact of these layouts typically rely on either computational fluid dynamics (CFD) or empirical analyses. However, these methods are often met with the constraints of computational intensity or diminished precision, particularly for complex module arrangements. 

This study presents an innovative approach that evaluates how the geometric layouts of PV modules affect their rate of convective heat loss under various environmental conditions. It achieves this by leveraging a combination of Physics-Informed Machine Learning (PIML) with a Deep Convolutional Neural Network (DCNN), to develop the PIML-DCNN approach. Additionally, we introduce a novel loss function denoted as `Pocket Loss', which ensures that the generated physical parameters within the PIML-DCNN model possess considerable interpretability by guiding the model to seek optimal results within a numerical range that aligns with physical principles. 

The developed PIML-DCNN model yields a relative error of 1.3\% on the testing dataset and an overall $R^2$ of 0.99 when estimating the averaged convective heat transfer coefficient of the PV array, as compared against validated CFD simulations. Once trained, the PIML-DCNN model can provide nearly instantaneous estimates of the convective heat transfer coefficient for new PV array configurations, striking an optimal balance between accuracy and computational efficiency. To delve into the influence of array configuration on power output, we examined the power output of each panel within a ten-row array. The analysis indicates that in arrays with staggered heights, the first three rows outperform those in uniformly heightened arrays in terms of power generation. However, from the fourth row onwards, a significant reduction in heat transfer efficiency becomes evident. Panels located in the downstream portion of staggered arrays suffer a more substantial efficiency loss than those in uniform configurations. Panels set at a uniform but elevated height across all rows demonstrate the most effective convective cooling performance.

In summary, the PIML-DCNN model that has been developed proves to be an effective tool for estimating the convective heat transfer capabilities of PV arrays across diverse configurations and environmental conditions. It offers critical insights into how array configurations affect convective heat transfer. As a result, this model presents considerable potential for refining the design of PV array configurations, ultimately enhancing the efficiency of PV power generation in real-world applications.

\section*{Appendix}
\subsection*{Validation of the CFD model}

\counterwithin{table}{section}
\renewcommand{\thetable}{A\arabic{table}}
\setcounter{table}{0}
\renewcommand{\thefigure}{A\arabic{figure}}
\setcounter{figure}{0}

\begin{figure}[!htbp]
\centering
    \includegraphics[width=1.0\textwidth]{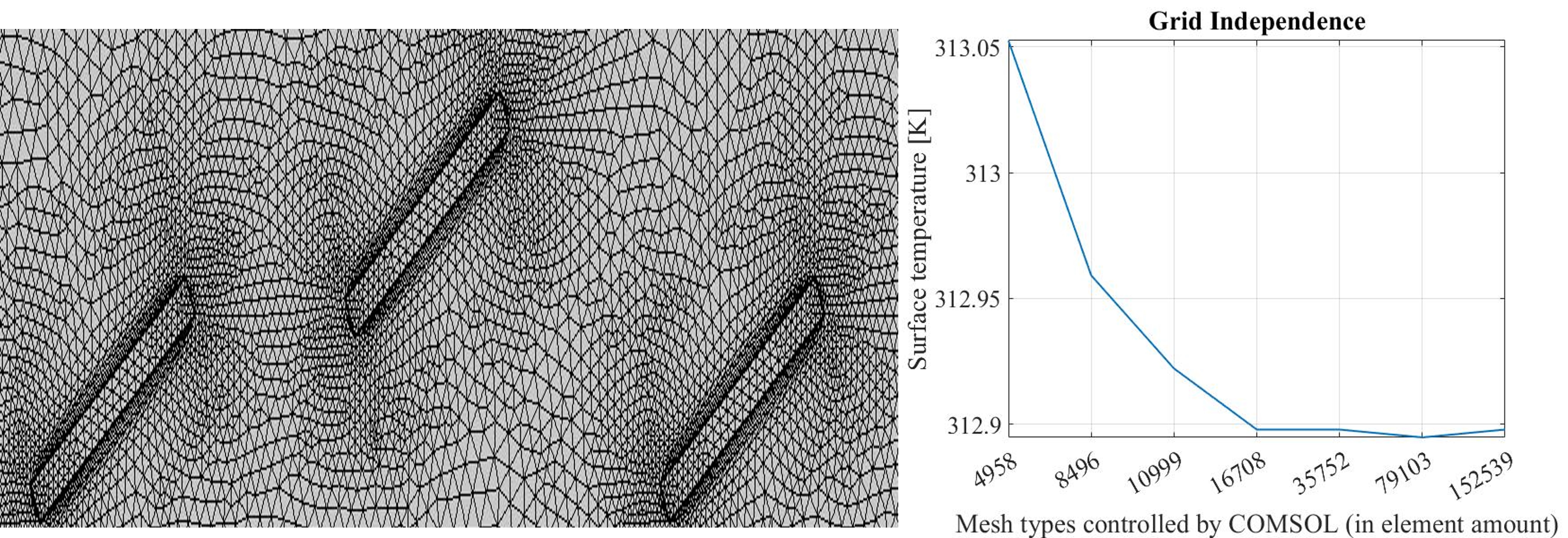}
\caption{Grid independence test of the LLL case with row spacing of 5.81 m and $U_{\infty}$ of 3.6 m/s.}
\label{fig:Grid}
\end{figure}

The grid independence test is illustrated in Fig.~\ref{fig:Grid}. When the number of elements in the mesh reaches 152,539, further refinement results in negligible differences in the simulated average surface temperature of the PV array. Consequently, this mesh density is utilized for all subsequent validation and numerical experiments.

The CFD results are validated against the Large Eddy Simulation (LES) results reported in~\cite{Stanislawski}. The LES model's fluid domain was three-dimensional, specifically configured to closely encompass the PV array. The spanwise length of this domain matches the width of a single PV panel with dimensional ratios of $W/L=0.01$ and $W/Z=0.04$. Considering that the spanwise length and the corresponding physical field variations are minimal relative to the other two dimensions, we suggest simplifying the model by omitting the spanwise dimension. This reduction aims to decrease computational costs while preserving an acceptable level of accuracy for heat transfer predictions.

For validation purposes, we simulate scenarios akin to those in~\cite{Stanislawski}, where panels held at a constant temperature of 320.15 K and a height of 1.52 m are spaced at varying intervals, under a wind speed $U_{\infty}$ of 3.6 m/s. The average convective heat transfer coefficient from the CFD simulations is computed using Eq.~(\ref{equation:governing equation}).



The Nusselt number, Nu, serves as the primary variable for validation. The relative difference $e$ is defined as:

\begin{equation}
e = \frac{|\text{Nu}_i - \widehat{\text{Nu}_i}|}{\min({\text{Nu}_i, \widehat{\text{Nu}_i}})} \times 100\%
\end{equation}
where $\text{Nu}_i$ is the Nusselt number derived from the benchmark LES, and $\widehat{\text{Nu}_i}$ represents the average Nusselt number obtained from our CFD model. For both methods, the panel height of 1.52 m is used as the characteristic length for calculating Nu in the validation scenarios. As shown in Table~\ref{Table: Validation Result}, the relative difference for each case stays within 8.3\%, affirming the reliability of our CFD model.



\begin{table}[!htbp]
\centering
\captionsetup{justification=centering}
\caption{Validation of CFD results.}
\begin{tabular}{cccc}
\hline
Row Spacing [m] & $ \text{Nu}_i$  & $\widehat{\text{Nu}_i}$ & $e$ \\ \hline
5.81 & 1735.6 & 1621.4        & 7.0\%        \\ 
6.54 & 1679.2 & 1650.3        & 1.7\%        \\ 
7.26 & 1621.2 & 1650.3        & 1.8\%        \\ 
7.99 & 1578.9 & 1664.7        & 5.4\%        \\ 
8.72 & 1551.0 & 1679.2        & 8.3\%        \\ \hline
\end{tabular}
\label{Table: Validation Result}
\vspace{1ex}
\end{table}

\subsection*{Proof of the universal approximator}\label{subsection:Approximator proof}

The linearization process within the PIML-DCNN model, specifically the feedforward propagation of Re without utilizing an activation function, can be succinctly described by the following set of equations, which embody the principles of a linear neural network:

\begin{equation}
\label{equation:1st layer}
y_1 = \mathbf{w}_1^\text{T} \text{Re} + b_1
\end{equation}

\begin{equation}
\label{equation:2nd layer}
\text{Nu} = \mathbf{w}_2^\text{T} y_1 + b_2
\end{equation}

Here, $\mathbf{w}_i$ and $b_i$ denote the weight matrix and bias vector of the $i^\text{th}$ layer~\cite{SVOZIL199743}, respectively. The first layer is designated as Dense4, and the second layer as Dense5. By combining Eqs.~(\ref{equation:1st layer}) and (\ref{equation:2nd layer}), we derive the comprehensive formula for computing the Nusselt number, Nu, from the Reynolds number, Re:

\begin{equation}
\label{equation:combined}
\text{Nu} = \mathbf{w}_2^\text{T}(\mathbf{w}_1^\text{T} \text{Re} + b_1) + b_2
\end{equation}

Upon simplification, since the term $\mathbf{w}_2^\text{T}b_1 + b_2$ is constant, the above equation effectively represents an approximation of a non-linear function of Re, as presented by Eq.(\ref{equation:approximator}). 

\section*{Acknowledgement}
This work is substantially supported by the Research Grants Council of the Hong Kong Special Administrative Region, China (Project No.~C6003-22Y). This work is also supported by the Hong Kong Polytechnic University Undergraduate Research and Innovation Scheme (Project No.~P0043659).

\section*{Declaration of Generative AI and AI-assisted technologies in the writing process}
During the preparation of this work, the authors used GPT-4 in order to improve the language. After using this tool/service, the authors reviewed and edited the content as needed and take full responsibility for the content of the publication.


\bibliographystyle{elsarticle-num-names} 

\biboptions{sort&compress} 

\bibliography{Final_ref}

\end{document}